\documentclass[aps, prd, letterpaper, twocolumn, amsmath, amssymb, preprintnumbers, superscriptaddress, floatfix, nofootinbib, longbibliography]{revtex4-1}
\usepackage[T1]{fontenc}
\usepackage{graphicx}
\usepackage{amsmath}
\usepackage{amsfonts}
\usepackage[colorlinks=true, linkcolor=blue, urlcolor=blue, citecolor=blue, anchorcolor=blue]{hyperref}

\newcommand{\al}{\ensuremath{\alpha} }
\newcommand{\be}{\ensuremath{\beta} }
\newcommand{\De}{\ensuremath{\Delta} }

\newcommand{\si}{\ensuremath{\sigma} }
\newcommand{\cO}{\ensuremath{\mathcal O} }
\newcommand{\psibar}{\ensuremath{\overline\psi} }
\newcommand{\MDM}{\ensuremath{M_{\text{DM}}} }
\newcommand{\lsim}{\ensuremath{\lesssim} }
\newcommand{\gsim}{\ensuremath{\gtrsim} }
\newcommand{\am}{\ensuremath{a\!\cdot\!m} }
\newcommand{\aNt}{\ensuremath{a\!\cdot\!N_t} }
\newcommand{\X}{\ensuremath{\!\times\!} }
\newcommand{\vev}[1]{\ensuremath{\left\langle #1 \right\rangle} }
\newcommand{\pbp}{\ensuremath{\vev{\psibar\psi}} }
\newcommand{\eq}[1]{Eq.~(\ref{#1})}
\newcommand{\fig}[1]{Fig.~\ref{#1}}
\newcommand{\tab}[1]{Table~\ref{#1}}
\newcommand{\refcite}[1]{Ref.~\cite{#1}}
\newcommand{\secref}[1]{Section~\ref{#1}}

\begin{document}
\title{Stealth dark matter confinement transition and gravitational waves}

\author{R.~C.~Brower}
\affiliation{Department of Physics and Center for Computational Science, Boston University, Boston, Massachusetts 02215, United States}
\author{K.~Cushman}
\affiliation{Department of Physics, Sloane Laboratory, Yale University, New Haven, Connecticut 06520, United States}
\author{G.~T.~Fleming}
\affiliation{Department of Physics, Sloane Laboratory, Yale University, New Haven, Connecticut 06520, United States}
\author{A.~Gasbarro}
\affiliation{AEC Institute for Theoretical Physics, University of Bern, 3012 Bern, Switzerland}
\author{A.~Hasenfratz}
\affiliation{Department of Physics, University of Colorado, Boulder, Colorado 80309, United States}
\author{X.~Y.~Jin}
\affiliation{Computational Science Division, Argonne National Laboratory, Argonne, Illinois 60439, United States}
\author{G.~D.~Kribs}
\affiliation{Department of Physics, University of Oregon, Eugene, Oregon, 97403 United States}
\author{E.~T.~Neil}
\affiliation{Department of Physics, University of Colorado, Boulder, Colorado 80309, United States}
\author{J.~C.~Osborn}
\affiliation{Computational Science Division, Argonne National Laboratory, Argonne, Illinois 60439, United States}
\author{C.~Rebbi}
\affiliation{Department of Physics and Center for Computational Science, Boston University, Boston, Massachusetts 02215, United States}
\author{E.~Rinaldi}
\affiliation{Arithmer Inc., R\&D Headquarters, Minato, Tokyo 106-6040, Japan}
\affiliation{Interdisciplinary Theoretical and Mathematical Sciences Program (iTHEMS), RIKEN, 2-1 Hirosawa, Wako, Saitama 351-0198, Japan}
\author{D.~Schaich}\email{david.schaich@liverpool.ac.uk}
\affiliation{Department of Mathematical Sciences, University of Liverpool, Liverpool L69 7ZL, United Kingdom}
\author{P.~Vranas}
\affiliation{Physical and Life Sciences, Lawrence Livermore National Laboratory, Livermore, California 94550, United States}
\affiliation{Nuclear Science Division, Lawrence Berkeley National Laboratory, Berkeley, California 94720, United States}
\author{O.~Witzel}
\affiliation{Department of Physics, University of Colorado, Boulder, Colorado 80309, United States}
\collaboration{Lattice Strong Dynamics Collaboration}
\noaffiliation

\preprint{LLNL-JRNL-811356}

\begin{abstract} 
  We use non-perturbative lattice calculations to investigate the finite-temperature confinement transition of stealth dark matter, focusing on the regime in which this early-universe transition is first order and would generate a stochastic background of gravitational waves.
  Stealth dark matter extends the standard model with a new strongly coupled SU(4) gauge sector with four massive fermions in the fundamental representation, producing a stable spin-0 `dark baryon' as a viable composite dark matter candidate.
  Future searches for stochastic gravitational waves will provide a new way to discover or constrain stealth dark matter, in addition to previously investigated direct-detection and collider experiments.
  As a first step to enabling this phenomenology, we determine how heavy the dark fermions need to be in order to produce a first-order stealth dark matter confinement transition.
\end{abstract}

\maketitle

\section{\label{sec:intro}Introduction and overview} 
The confining gauge--fermion theory of quantum chromodynamics (QCD) produces the massive stable protons and nuclei of the visible universe, making it compelling to hypothesize that new strong dynamics could also underlie the dark sector.
Stealth dark matter~\cite{Appelquist:2015yfa, Appelquist:2015zfa} is a particularly attractive model of composite dark matter, based on a new strongly interacting SU($N_D$) gauge sector with even $N_D \geq 4$, which is coupled to four massive fermions in the fundamental representation.
As detailed in \refcite{Appelquist:2015yfa}, the four `dark fermions' transform in non-trivial vector-like representations of the electroweak group, in order to generate the correct cosmological dark matter abundance while also satisfying all experimental constraints.
Although these `dark' fermions are electrically charged and couple to the standard model (SM) Higgs boson, following the dark-sector confinement transition they give rise to a composite dark matter candidate in the form of the lightest spin-0 SU($N_D$) `dark baryon', which is a singlet under the entire SM gauge group.
This dark matter candidate is automatically stable on cosmological time scales due to the conservation of dark baryon number, and it acquires mass both from confinement and from the masses of its fermion constituents.

Experimental constraints on the stealth dark matter model come from both direct-detection searches and collider experiments, with direct-detection cross sections arising from non-perturbative form factors of the dark baryon.
For example, direct detection through Higgs boson exchange depends on the dark baryon's scalar form factor, as well as on the relative sizes of the vector-like and electroweak-breaking fermion mass terms that appear in the model's lagrangian~\cite{Appelquist:2015yfa}.
Existing direct-detection searches, combined with lattice calculations of that scalar form factor, require that the vector-like contributions to the dark fermions' masses dominate over the electroweak-breaking contributions~\cite{Appelquist:2015yfa, Appelquist:2014jch}.
Those lattice calculations considered the minimal case $N_D = 4$, which is also the case we will consider in this work.
This choice minimizes the computational costs of our lattice calculations, while still being large enough for large-$N$ scaling relations to recast results to larger $N_D \geq 6$ with reasonable reliability.
(See \refcite{Lucini:2012gg} for a thorough review of the large-$N$ framework.)

Direct detection can also proceed through photon exchange, and the symmetries of the model strongly suppress this cross section by forbidding the leading magnetic moment and charge radius contributions to it.
The contribution from the dark baryon's electromagnetic polarizability is unavoidable, and provides a lower bound on direct-detection signals for the entire class of dark matter models featuring neutral dark baryons with charged constituents (reviewed in \refcite{Kribs:2016cew}).
Lattice calculations of that polarizability~\cite{Appelquist:2015zfa}, again for the case $N_D = 4$, obtain the constraint $\MDM \gsim 0.2$~TeV from existing direct-detection searches.\footnote{This SU(4) result can be contrasted with the direct-detection constraint $\MDM \gsim 20$~TeV for an SU(3) model with unsuppressed magnetic moment and charge radius interactions~\cite{Appelquist:2013ms}.}
The steep dependence of the cross section on the dark baryon mass, $\si \propto 1 / \MDM^6$, causes the predicted signal to fall below the irreducible neutrino background for $\MDM \gsim 0.7$~TeV~\cite{Appelquist:2015zfa}.

Stronger constraints on stealth dark matter currently come from collider searches for vector ($V$) and pseudoscalar ($P$) `dark mesons', some of which are electrically charged.
If $M_P / M_V < 0.5$ so that $V \to PP$ decay is possible, the dark vector meson becomes a broad resonance and masses as light as $M_P \simeq 0.13$~TeV and $M_V \simeq 0.3$~TeV remain viable~\cite{Kribs:2018ilo}.
Lattice calculations of the meson and baryon spectrum can translate these bounds into constraints on $\MDM > M_V$.
In this work we will focus on the heavy-mass regime, $M_P / M_V > 0.5$, where $V \to PP$ decays are kinematically forbidden.
The dominant decay process is then $V \to \ell^+ \ell^-$, which could be observed in searches for $Z' \to \ell^+ \ell^-$.
This produces the constraint $M_V \gsim 2$~TeV reported by \refcite{Kribs:2018ilo}, assuming this process is dominated by a single dark vector meson. 
In the heavy-mass regime, we can approximate $\MDM \simeq \frac{N_D}{2} M_V$ to turn this into a lower bound on the dark baryon mass.

It is difficult to set an upper bound on the mass of the dark baryon, though some very rough estimates can be made by requiring that the stealth dark matter model produces the observed cosmological dark matter abundance.
Specifically, \refcite{Appelquist:2015yfa} estimates that a predominantly symmetric thermal abundance of stealth dark matter would match cosmology for \MDM of order tens to hundreds of TeV, while \MDM smaller than a few TeV would require a predominantly asymmetric abundance.
There is therefore a significant allowed range of stealth dark matter masses up to hundreds of TeV, which will be very challenging for direct detection or collider experiments to constrain.

This makes the possibility of using gravitational waves to constrain or discover stealth dark matter particularly exciting.
There is increasing interest in probing dark sectors by searching for a stochastic background of gravitational waves that would be produced by a first-order phase transition in the early universe~\cite{Schwaller:2015tja, Jaeckel:2016jlh, Huang:2017rzf, Aoki:2017aws, Huang:2017kzu, Croon:2018erz, Mazumdar:2018dfl, Christensen:2018iqi, Breitbach:2018ddu, Baratella:2018pxi, Fairbairn:2019xog, Helmboldt:2019pan, Bertone:2019irm, Archer-Smith:2019gzq, Aoki:2019mlt}. 
Such searches are an important component of the science programs for future space-based facilities including the LISA observatory~\cite{Caprini:2015zlo, Caprini:2019egz}, DECIGO~\cite{Sato:2017dkf} and AEDGE~\cite{Bertoldi:2019tck}.
This approach has the advantage of involving only gravity, the force that provides the existing astrophysical and cosmological evidence for dark matter.
In the context of strongly coupled composite models such as stealth dark matter, the transition of interest is the confinement transition through which the state of the system changes from a high-temperature deconfined plasma of `dark gluons' and dark fermions to stable SM-singlet dark baryons.
If this confinement transition was first order, its properties including the nucleation temperature and latent heat govern the stochastic spectrum of the gravitational waves it produced, making reliable knowledge of these properties a crucial ingredient to extract constraints from future observations~\cite{Schwaller:2015tja, Caprini:2015zlo, Caprini:2019egz}.

\begin{figure}[tbp]
  \includegraphics[width=\linewidth]{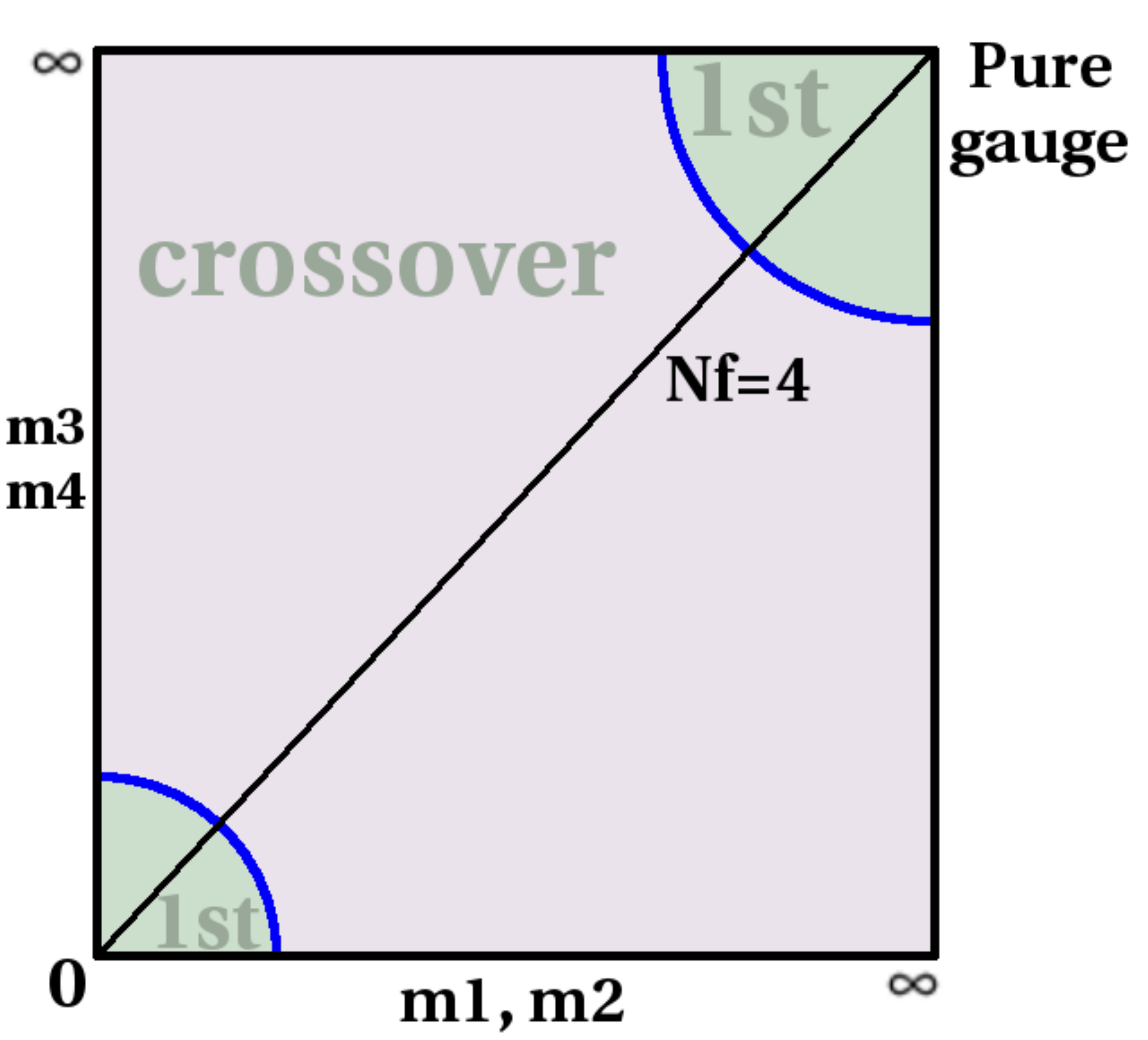}
  \caption{\label{fig:columbia}A sketch of the `Columbia plot' for SU($N$) gauge theories coupled to two pairs of fundamental fermions, taking $N \geq 3$ so that the confinement transition is first order when all four fermions are either sufficiently heavy or sufficiently light.}
\end{figure}

In this paper we use non-perturbative lattice calculations to investigate the finite-temperature confinement transition of SU(4) stealth dark matter.
We focus on the first goal of determining the region of parameter space for which the confinement transition of this gauge--fermion theory is first order, in contrast to the continuous crossover of QCD.
Achieving this first goal is a necessary step to enable more detailed future lattice investigations of the resulting gravitational waves.
Some preliminary results from this work previously appeared in \refcite{Schaich:2020vaj}.

The `Columbia plot'~\cite{Brown:1990ev} shown in \fig{fig:columbia} illustrates what we can expect based on symmetry arguments and continuum effective models~\cite{Yaffe:1982qf, Pisarski:1983ms, Svetitsky:1985ye}.
Although we specialize this version of the plot to the two pairs of degenerate fermions that stealth dark matter involves, for generic SU($N$) gauge theories with $N \geq 3$ and $N_f \lsim 2N$ fundamental fermions, first-order transitions are expected in two regimes: where the fermions are sufficiently heavy or sufficiently light.
These expectations have been supported by lattice calculations, though at present controlled continuum extrapolations have been achieved by lattice analyses of only two points on the Columbia plot.
One of these is the (2+1)-flavor physical point of SU(3) QCD---see the recent review \refcite{Philipsen:2019rjq} and references therein.
The other is the SU($N$) pure-gauge system that corresponds to the infinite-mass limit in the upper-right corner of the plot, for $3 \leq N \leq 10$~\cite{Gavai:2002td, Lucini:2005vg, Datta:2009jn, Datta:2010sq, Lucini:2012gg}.

In this work we will focus on the SU(4) heavy-mass first-order transition region connected to the pure-gauge limit.
Compared to the light-mass region, this both reduces computational costs and limits the reach of collider constraints on $\MDM$, which become more powerful as the ratio $\MDM / M_P$ grows towards the $M_P \to 0$ chiral limit.
Although stealth dark matter requires at least a small mass splitting between the two pairs of degenerate fermions, in order to guarantee that all `dark mesons' are unstable and do not disrupt Big Bang nucleosynthesis~\cite{Appelquist:2015yfa}, for simplicity we will consider in this work only the limit of four degenerate flavors, corresponding to the ``$N_f = 4$'' diagonal line in \fig{fig:columbia}.
In \secref{sec:conc} we will discuss prospects for future investigations of the more general non-degenerate situation.

The first goal mentioned above is now a matter of determining how heavy these $N_f = 4$ dark fermions need to be in order to produce a first-order stealth dark matter confinement transition.
This investigation is the first lattice study of the heavy-mass region of the Columbia plot for SU(4) gauge theory with dynamical fermions.
Even for the case of SU(3) this region has received relatively little attention compared to the QCD physical point and chiral limit.
See \refcite{Ejiri:2019csa} (and references therein) for a recent SU(3) investigation, which finds that very large masses are needed to produce a first-order transition.
These masses are parameterized by the ratio $M_P / T_c \gsim 10$, where $T_c$ is the equilibrium critical temperature and the extrapolation to the continuum limit is not yet under control.
The need for very large masses for a first-order SU(3) transition may be related to the known weakness of the first-order confinement transition in the SU(3) pure-gauge limit~\cite{Lucini:2012gg, Ejiri:2019csa}.
Since this pure-gauge confinement transition strengthens significantly with increasing $N \geq 4$~\cite{Lucini:2005vg, Datta:2009jn, Lucini:2012gg}, stealth dark matter may exhibit qualitatively different behavior, motivating our dedicated lattice calculations.

We begin in the next section by explaining the strategy of our lattice calculations, including our nHYP-improved unrooted-staggered lattice action, the SU(4) ensembles we have generated using it, and the observables we focus on to analyze the confinement transition.
Through \refcite{data} we provide a comprehensive data release summarizing our ensembles and results.
In \secref{sec:0f} we test our methods by considering the pure-gauge limit, which provides a less-expensive means to assess the discretization artifacts of our lattice action.
We also exploit our prior knowledge that the pure-gauge SU(4) transition is strongly first-order, which allows us to view our pure-gauge results as a guide to the signals we should expect for a first-order confinement transition with dynamical fermions.
In \secref{sec:4f} we add those $N_f = 4$ degenerate dynamical fermions, and supplement our finite-temperature analyses with zero-temperature meson spectroscopy calculations.
These ingredients allow us to determine the ratio of dark pion and dark vector meson masses, $M_P / M_V > 0.9$, required for the stealth dark matter confinement transition to be first order.

We discuss our conclusions in \secref{sec:conc}, and look ahead to our follow-up work that will investigate this first-order transition in more detail, in order to predict more detailed features of the gravitational waves it would produce.
Key parameters that need to be computed or estimated to predict the gravitational-wave spectrum are the latent heat (or vacuum energy fraction), the phase transition duration, and the bubble wall velocity~\cite{Kamionkowski:1993fg, Espinosa:2010hh}.
Only the first of these is straightforward to determine through lattice calculations, and this will be the next focus of our investigations.
Even without a careful continuum-extrapolated analysis of the latent heat, our results reported in this paper will allow future searches for stochastic gravitational waves (resulting in either detections or exclusions) to set novel constraints on stealth dark matter and similar models.
For example, the gravitational-wave spectrum also depends on the transition temperature $T_*$, which may differ from the equilibrium critical temperature $T_c$ used to set the scale of our lattice calculations, due to possible supercooling.
If we can assume $T_* \simeq T_c$ or estimate how they differ, then our results for the mass dependence of the stealth dark matter transition will translate information on $T_*$ from gravitational-wave searches into predictions for both the approximate mass scale of the dark baryons as well as the minimum masses of the dark mesons being searched for at colliders.

\section{\label{sec:lattice}Lattice setup and strategy} 
\subsection{Context and lattice action} 
As usual~\cite{DeTar:2009ef}, our SU(4) lattice calculations employ a hypercubic grid of $L^3\X N_t$ sites defining a discrete euclidean space-time.
We impose thermal boundary conditions (periodic for bosons, antiperiodic for fermions) in the temporal direction, while all fields are subject to periodic boundary conditions in the spatial directions.
The lattice spacing `$a$' between neighboring lattice sites is set through the input bare gauge coupling $\be_F \propto 1 / g_0^2$, which we discuss in more detail below.
The temperature in `lattice units' is the inverse temporal extent of the lattice, $T = 1 / (\aNt)$, and in the finite-temperature context we are interested in $N_t < L$.

For a fixed lattice volume $L^3\X N_t$ we proceed by varying the bare coupling $\be_F$ to scan the temperature.
Below we discuss the observables we monitor as functions of the coupling, which reveal the critical $\be_F^{(c)}$ corresponding to $T_c$, and provide information about the order of the transition.
Setting the lattice scale by taking $T_c = 1 / (a_c\!\cdot\!N_t)$ to be a fixed physical temperature means that the lattice spacing at the transition decreases as $N_t$ increases, identifying the $a \to 0$ continuum limit with the limit $N_t \to \infty$.
If $N_t$ is too small, the large lattice spacing may result in significant systematic errors from discretization artifacts.

At the same time, the aspect ratio $L / N_t$ must be sufficiently large to ensure that systematic errors from the finite spatial volume are also under control.
This motivates keeping $N_t$ as small as discretization artifacts allow.
The large lattice spacings at small-$N_t$ thermal transitions correspond to strong bare gauge couplings, and studies spanning many years~\cite{Bhanot:1981eb, Lucini:2013wsa} have observed that such strong couplings for can produce a bulk (zero-temperature) transition into a lattice phase with no continuum limit.
For SU(4) Yang--Mills theory with a lattice action that includes both fundamental and adjoint plaquette terms, with respective couplings $\be_F$ and $\be_A$, this bulk transition is first order for sufficiently large $\be_A > 0$, with a cross-over persisting when $\be_A = 0$.\footnote{For SU($N$) Yang--Mills theories with $N \geq 5$, the first-order bulk transition extends into the $\be_A < 0$ regime~\cite{Lucini:2005vg, Lucini:2012gg}.  Recall that the plaquette ($\Box$) in a given representation is the gauge-invariant trace in that representation of the product of gauge links around an elementary face of the lattice.}
With $N_t \lsim 4$, thermal transitions for $\be_A = 0$ effectively merge with this bulk crossover, resulting in unmanageable discretization artifacts.

In an attempt to ameliorate this problem, we follow \refcite{Cheng:2011ic} and use a negative adjoint coupling $\be_A = -\be_F / 4$ in the fundamental--adjoint gauge action.
At tree level
\begin{equation}
  \label{eq:FA}
  \frac{2N}{g_0^2} = \be_F + 2\be_A = \be_F \left(1 + 2\frac{\be_A}{\be_F}\right)
\end{equation}
for SU($N$) gauge theory, requiring $\be_A > -\be_F / 2$.
This tree-level relation is not accurate at the critical $\be_F^{(c)}$ of the thermal transitions with $N_t \leq 12$, which can be seen by contrasting our pure-gauge results in the next section against past studies of SU(4) lattice gauge theory using $\be_A = 0$~\cite{Wingate:2000bb, Lucini:2005vg, Panero:2009tv, Datta:2009jn, Datta:2010sq}.
Continuum extrapolations would therefore be required to quantitatively compare our pure-gauge results (e.g., for the latent heat) with that earlier work.
The same is true for comparisons with \refcite{Gavai:2002td}, which avoids strong-coupling bulk transitions by modifying the lattice action to restrict the fundamental plaquette to a single $Z_4$ vacuum.

Unlike those prior pure-gauge studies, we also carry out calculations with four dynamical fermions in the fundamental representation of SU(4).
As discussed in \secref{sec:intro}, for simplicity we consider only four degenerate flavors, which allows us to use an unrooted staggered-fermion lattice action.
To reduce discretization artifacts for the relatively large fermion masses \am that we will consider, we also improve the fermion action by incorporating smearing.
Again following \refcite{Cheng:2011ic}, we use a single nHYP smearing step~\cite{Hasenfratz:2001hp, Hasenfratz:2007rf} with parameters $(0.5, 0.5, 0.4)$.

\subsection{Strategy} 
The considerations above lead us to the following strategy for the ensembles of gauge configurations we generate to map out the finite-temperature SU(4) phase diagram.
\begin{itemize}
  \item We need to consider several fermion masses \am in order to determine the regime in which the stealth dark matter confinement transition is first order.\footnote{A second-order transition is expected for the critical value of \am at the endpoint separating the line of first-order transitions from the continuous crossover at smaller masses.  While the masses we consider are unlikely to land precisely on this critical point, its proximity could influence the transition signals discussed below.}
        Our smallest fermion mass $\am = 0.05$ is chosen to overlap the mass range considered in Refs.~\cite{Appelquist:2015yfa, Appelquist:2015zfa}.
        We also carry out pure-gauge calculations corresponding to the $\am \to \infty$ quenched limit.
        In total we consider $\am = \left\{0.05, 0.1, 0.2, 0.4, \infty\right\}$.
  \item For each of those five $\am$, we want at least three $N_t$ in order to enable $N_t \to \infty$ continuum extrapolations.
        In total we consider $N_t = \left\{4, 6, 8, 12\right\}$, but we will see in the next section that $N_t = 4$ may suffer from large discretization artifacts despite our improved lattice action.
        We will therefore use $N_t \geq 6$ to carry out continuum extrapolations, which remain work in progress.
        While these continuum extrapolations will be important for our subsequent studies of (e.g.) the latent heat, they are not crucial for our present task of determining the SU(4) phase diagram.
        In this work we will focus on $N_t =  8$, the largest temporal extent for which large amounts of data are available, using the other $N_t$ primarily to assess discretization artifacts.
  \item For each $\left\{\am, N_t\right\}$ we want at least three aspect ratios $L / N_t \geq 2$ in order to enable extrapolations to the thermodynamic limit of infinite spatial volume.
        In our present work these multiple spatial volumes are most useful for distinguishing between first-order transitions and continuous crossovers, for instance from the $L$ dependence of relevant susceptibilities or kurtoses.
        More careful infinite-volume extrapolations will again feature in our upcoming detailed studies of transition properties.
        So far we have considered aspect ratios $L / N_t = \left\{2, 3, 4, 6, 8\right\}$.
  \item Finally, for each $\left\{\am, N_t, L / N_t\right\}$, we scan in temperature by varying the input bare fundamental coupling $\be_F$.
        We begin at a high value of $\be_F$ deep in the deconfined phase and systematically lower the temperature through the transition and into the confined phase, starting each lower-temperature calculation from a thermalized gauge-field configuration generated at slightly higher $\be_F$.
        Once we are deep in the confined phase we reverse this process and also scan from low to high temperature in order to check for possible hysteresis.
        Following these initial coarse scans with relatively large $\De \be_F = 1$--$2$ between subsequent calculations, we carry out one or two rounds of refined scans around the transition region with smaller $0.02 \leq \De \be_F \leq 0.2$.
\end{itemize}

For both pure-gauge and dynamical calculations we use the hybrid Monte Carlo (HMC) algorithm~\cite{Duane:1987de}, employing QHMC/FUEL~\cite{Osborn:2014kda} on top of the USQCD SciDAC software stack,\footnote{\texttt{\href{http://usqcd-software.github.io}{usqcd-software.github.io}}} which provides efficient performance for arbitrary SU($N$) gauge groups.
We use a second-order Omelyan integrator~\cite{Takaishi:2005tz} with multiple time scales~\cite{Urbach:2005ji} and (for $\am < \infty$) an additional heavy pseudofermion field~\cite{Hasenbusch:2002ai}, fixing a trajectory length of $\tau_{\text{traj}} = 1$ molecular dynamics time unit (MDTU) and tuning molecular dynamics step sizes to target roughly 60\%--80\% acceptance rates~\cite{Takaishi:2005tz}.
We monitor the `Creutz equality'~\cite{Creutz:1988wv} $\vev{e^{-\De H}} = 1$ to ensure that our HMC parameter choices are appropriate. 
We also accumulate a similar number of MDTU for both pure-gauge and dynamical calculations.
While larger volumes and higher statistics could be obtained with more efficient algorithms in the pure-gauge case, our goal here is to use this known first-order transition to illuminate the signal quality we may expect from the algorithms and statistics available to us in the more expensive dynamical case.

In total, with $\am = \left\{0.05, 0.1, 0.2, 0.4, \infty\right\}$, $N_t = \left\{4, 6, 8, 12\right\}$ and $L / N_t = \left\{2, 3, 4, 6, 8\right\}$ we have generated 1,381 finite-temperature HMC Markov chains (or `streams'), each with at least 2,000 MDTU and up to 75,000 MDTU.
We use the same HMC parameters for both high- and low-start streams, which allows us to combine 1,166 of these streams into 583 joint ensembles with approximately doubled statistics.
Table~\ref{tab:ensembles} summarizes these streams and their organization.
In addition, we generated 12 zero-temperature ensembles with lattice volume $24^3\X 48$, at the critical coupling $\be_F^{(c)}$ and at $\be_F^{(c)} \pm 0.2$ for each $\am < \infty$.
We use these zero-temperature ensembles to compute the meson spectrum and relate \am to the ratio of dark pion and dark vector meson masses, $M_P / M_V$.
This provides a convenient parameterization of the fermion masses that can easily be compared to previous quenched lattice studies of stealth dark matter~\cite{Appelquist:2015yfa, Appelquist:2015zfa}, which used valence Wilson fermions with $0.55 \lsim M_P / M_V \lsim 0.77$.

The variation in the number of MDTU per finite-temperature stream is driven by auto-correlations that increase significantly around the transition (even if the `transition' is a continuous crossover), requiring longer HMC streams in this region.
For each stream we set a thermalization cut by hand based on human inspection of time-series plots, and use the `autocorr' module in \texttt{emcee}~\cite{Foreman:2013mc} to estimate auto-correlation times $\tau$ for selected non-topological observables discussed below.
We then divide our measurements into bins for jackknife analyses, with bin sizes larger than $\tau$ and at least 100~MDTU, collecting sufficient data to ensure that at least ten such statistically independent bins are available.
The maximum auto-correlation time we observe, $\tau \approx 4750$~MDTU, produces 26 jackknife bins, 13 from each of the high- and low-start streams.
All of these details and many more are provided through our data release \refcite{data}.

\subsection{\label{sec:obs}Observables} 
The key observable signalling the confinement transition is the Polyakov loop ($PL$), the gauge-invariant trace of the product of gauge links wrapping around the temporal extent of the lattice.
In the pure-gauge SU($N$) theory, the Polyakov loop is an order parameter of the (temporal) $Z_N$ center symmetry, which breaks spontaneously in the high-temperature deconfined phase where the magnitude $|PL| \to N$ as $\be_F \to \infty$ and the argument is restricted to lie near any one of the $N$ degenerate vacua oriented at $e^{i\phi} = e^{2\pi i k / N}$ with $k = 0, \cdots, N - 1$.
Dynamical fermions in the fundamental representation explicitly break this center symmetry, picking out the positive real axis ($\phi = 0$) as the preferred vacuum.
In order to apply identical analyses to both the pure-gauge and dynamical theories, we focus on the magnitude $|PL|$ as the most useful observable.

We improve the signal for the Polyakov loop by computing it after smoothing the lattice gauge fields by applying the Wilson flow, a continuous transformation that systematically removes short-distance lattice cutoff effects~\cite{Narayanan:2006rf, Luscher:2009eq}.
This Wilson-flowed Polyakov loop $PL_W$ is a modern variant of the RG-blocked Polyakov loop investigated in older works~\cite{Schaich:2012fr, Hasenfratz:2013uha}, and has previously been used in Refs.~\cite{Schaich:2015psa, Datta:2015bzm, Wandelt:2016oym, Ayyar:2018ppa, Appelquist:2018yqe}.
The removal of short-distance fluctuations significantly enhances the signal without affecting the physics of the transition, producing much clearer contrasts between confined systems with small $|PL_W| \ll 1$ and deconfined systems with large $|PL_W| \sim N$.
We restrict the `flow time' $t$ by requiring $c \equiv \sqrt{8t} / N_t \leq 0.5$ or equivalently $t \leq N_t^2 / 32$.
Since $4 \leq N_t \leq 12$, this maximal $c = 0.5$ still corresponds to modest flow times $0.5 \leq t \leq 4.5$, respectively.
In this paper we will therefore only show results obtained with $c = 0.5$.
Behind the scenes we also monitor $c = 0.2$, $0.3$ and $0.4$ to check that our focus on $c = 0.5$ doesn't introduce systematic errors.
In particular, $|PL_W|$ with $c = 0.5$ is the main observable whose auto-correlation time we monitor to set jackknife bin sizes.\footnote{We also monitor the auto-correlation time of the chiral condensate $\pbp$, but the relatively large masses we consider strongly break chiral symmetry and leave \pbp of little use for analyzing the confinement transition.}

In addition to the expectation value $\vev{|PL_W|}$ itself, we also compute the susceptibility
\begin{equation}
  \chi_{\cO} = L^3 \left(\vev{\cO^2} - \vev{\cO}^2\right)
\end{equation}
and kurtosis (equivalent to the Binder cumulant)
\begin{equation}
  \label{eq:kurtosis}
  \kappa_{\cO} = \frac{\vev{\cO^4} - 4\vev{\cO^3}\vev{\cO} + 6\vev{\cO^2}\vev{\cO}^2 - 3\vev{\cO}^4}{\chi_{\cO}^2}
\end{equation}
for the (volume-averaged) Wilson-flowed Polyakov loop magnitude $\cO = |PL_W|$.
This susceptibility exhibits a peak at the confinement transition, with the order of the transition reflected by the $L$-dependence of the peak height and of the kurtosis~\cite{Kuramashi:2020meg}.
We will similarly use the plaquette susceptibility $\chi_{\Box} = L^3 N_t \left(\vev{\Box^2} - \vev{\Box}^2\right)$ to identify the zero-temperature bulk phase transition.
Because the plaquette is much less noisy than the Polyakov loop, there is no need to improve its signal with the Wilson flow.

Another quantity sensitive to the confinement transition is the spatial/temporal anisotropy of the Wilson-flowed energy density $t^2\vev{E(t)}$~\cite{Datta:2015bzm, Wandelt:2016oym, Ayyar:2018ppa} (which was initially considered by \refcite{Borsanyi:2012zr} in the context of tuning anisotropic lattice spacings).
Following \refcite{Ayyar:2018ppa} we analyze the ratio
\begin{equation}
  R_E(t) \equiv \vev{\frac{E_{ss}(t)}{E_{s\tau}(t)}},
\end{equation}
where the `space--space' $E_{ss}(t)$ is computed from `clover' terms built out of four plaquettes oriented in the purely spatial planes $x$--$y$, $x$--$z$ and $y$--$z$, while the clover terms contributing to the `space--time' $E_{s\tau}(t)$ are oriented in the $x$--$\tau$, $y$--$\tau$ and $z$--$\tau$ planes.
We will again focus on values of the flow time $t$ corresponding to $c = 0.5$.
In the low-temperature confined phase, the system is isotropic and $R_E \approx 1$, while the breaking of temporal (but not spatial) center symmetry in the high-temperature deconfined phase produces $R_E > 1$.

Finally, we also monitor the `deconfinement fraction' discussed in Refs.~\cite{Wingate:2000bb, Christ:1985wx}, which measures the proportion of Polyakov loop measurements whose arguments fall within a certain (tunable) angle $\theta < \pi / 4$ around any of the $Z_4$ vacua.
As above, we consider the Wilson-flowed $\arg(PL_W)$ at flow times corresponding to $c = 0.5$.
With $N_{\text{in}}$ of $N_{\text{tot}}$ measurements suitably aligned along the $Z_4$ axes, we define the deconfinement fraction
\begin{equation}
  \label{eq:deconf_frac}
  f(\theta) \equiv \frac{\pi / 4}{\pi / 4 - \theta}\left[\frac{N_{\text{in}}}{N_{\text{tot}}} - \frac{\theta}{\pi / 4}\right]
\end{equation}
so that $f \to 1$ in the deconfined phase where $N_{\text{in}} \approx N_{\text{tot}}$, and $f \to 0$ in the confined phase where $\arg(PL_W)$ is approximately uniformly distributed in $[0, 2\pi)$.
While this quantity was originally developed in the context of pure-gauge theories, it remains well-defined in the presence of dynamical fermions.
Results for $f(\theta)$ depend on the tunable parameter $\theta$, and we make the assumption that the systematic effects of choosing $\theta$ dominate the total uncertainty in the deconfinement fraction.
Computing the central value with $\theta = 0.2 \approx 11.5^{\circ}$, we therefore set the uncertainty on $f(\theta)$ by varying $\theta \in [0.15, 0.25] \approx [8.6^{\circ}, 14.3^{\circ}]$.

Using these observables, we will now reproduce the well-studied first-order confinement transition in pure-gauge SU(4) Yang--Mills theory, and use that experience to investigate the mass dependence of the stealth dark matter confinement transition with $N_f = 4$ degenerate dynamical fermions.

\section{\label{sec:0f}Pure-gauge limit} 
Over the years there have been several lattice investigations of the SU($N$) Yang--Mills confinement transition with $N > 3$, primarily exploring the approach to the large-$N$ limit.
See Refs.~\cite{Wingate:2000bb, Gavai:2002td, Lucini:2005vg, Panero:2009tv, Datta:2009jn, Datta:2010sq} for work with a focus on $N = 4$ (building on much earlier studies~\cite{Gocksch:1984mg, Green:1984ks, Batrouni:1984vd, Wheater:1984wd}) and \refcite{Lucini:2012gg} for a broader review. 
We revisit this calculation with two main goals, in addition to confirming that our code and algorithms are working correctly.
First, we will use the computationally inexpensive pure-gauge limit to check the discretization artifacts of our improved fundamental--adjoint gauge action, and assess which $N_t$ will be safe to use in dynamical calculations without complications from the bulk transition discussed above.
Second, our prior knowledge that the pure-gauge SU(4) transition is strongly first-order allows us to observe the quality of signals we should expect for a first-order transition with dynamical fermions, which will be useful to distinguish this case from a continuous crossover in \secref{sec:4f}.

\subsection{Discretization artifacts} 
\begin{figure}[tbp]
  \includegraphics[width=\linewidth]{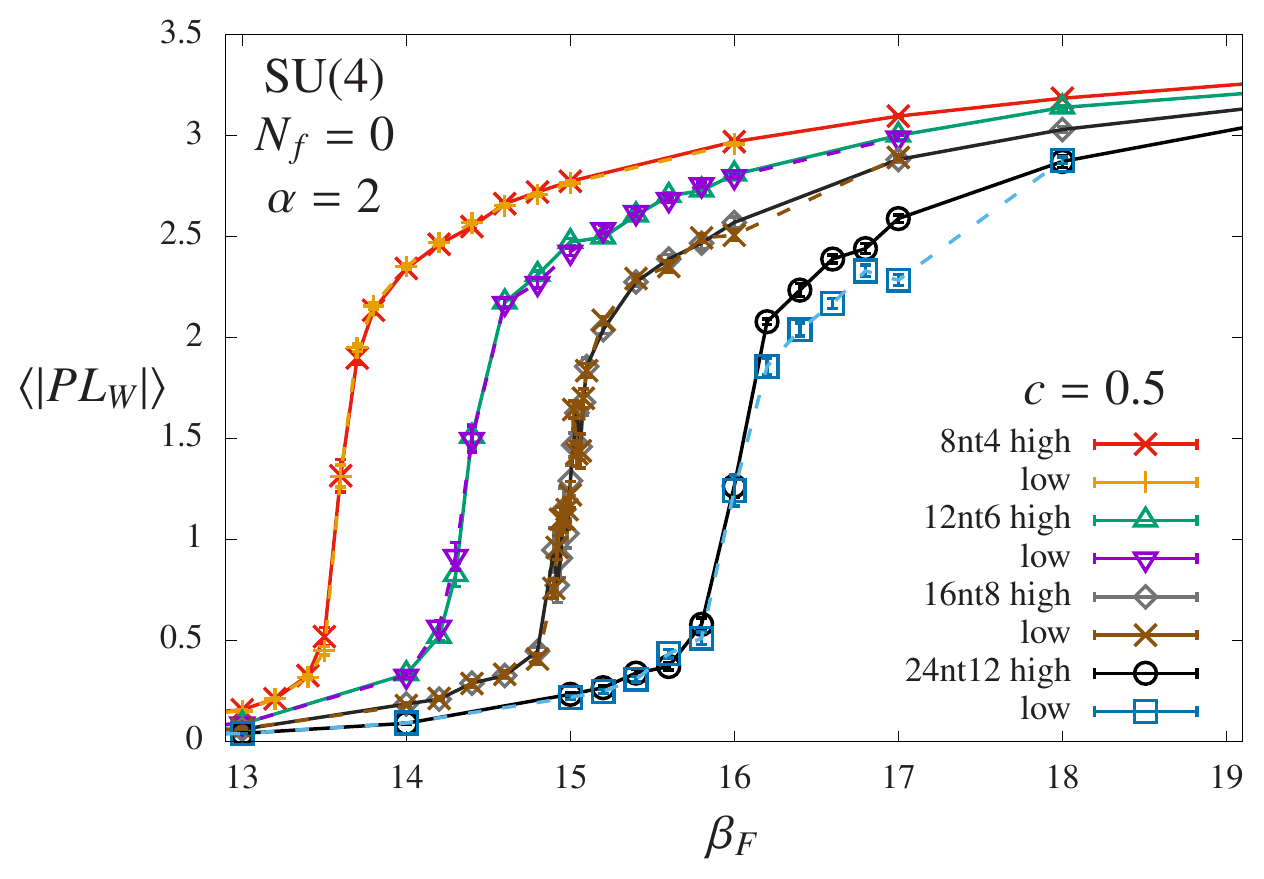}\\[12 pt]
  \includegraphics[width=\linewidth]{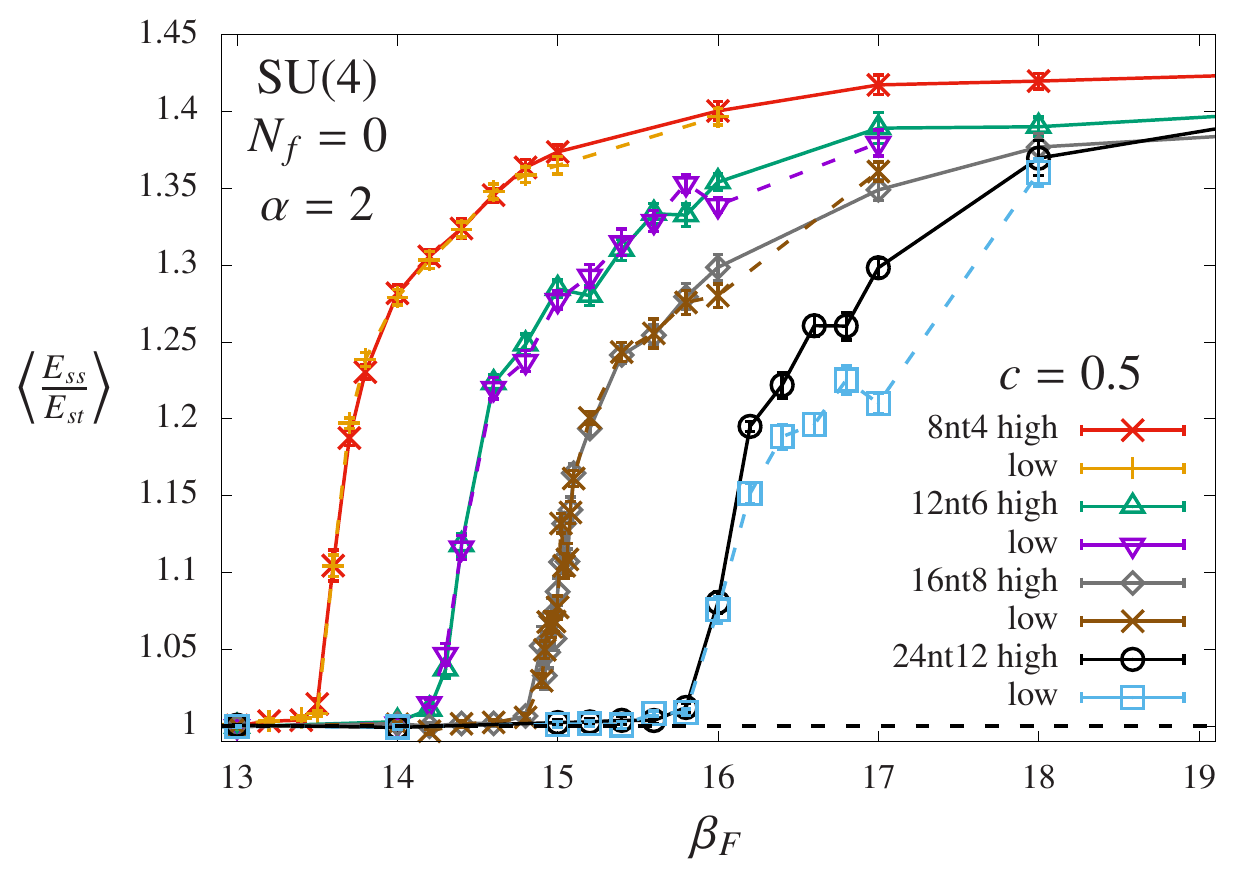}
  \caption{\label{fig:Wpoly_Nt}Dependence of the pure-gauge SU(4) critical coupling $\be_F^{(c)}$ on the temporal extent of the lattice $N_t$, comparing lattice volumes $8^3\X 4$, $12^3\X 6$, $16^3\X 8$ and $24^3\X 12$ with aspect ratio $\al \equiv L / N_t = 2$.  As $N_t$ decreases, confinement occurs at stronger couplings (smaller $\be_F^{(c)}$), as shown by both the Wilson-flowed Polyakov loop magnitude $|PL_W|$ (top) and the Wilson-flowed $E(t)$ anisotropy (bottom).  We plot separate results for the high- and low-start streams, with lines connecting points to guide the eye, to show the absence of hysteresis.  The small differences between the high- and low-start $N_t = 12$ results for $16.2 \leq \be_F \leq 17$ are discussed in the text.}
\end{figure}

In \fig{fig:Wpoly_Nt} we show how the critical coupling $\be_F^{(c)}$ of the pure-gauge thermal confinement transition depends on the temporal extent of the lattice $N_t$, to clarify the more abstract discussions in \secref{sec:lattice} above.
With fixed aspect ratio $L / N_t = 2$ for $N_t = 4$, 6, 8 and 12, the transition is clear in both the Wilson-flowed Polyakov loop magnitude $|PL_W|$ and the Wilson-flowed $E(t)$ anisotropy, illustrating the behavior described in \secref{sec:lattice}.
As $N_t$ increases, the fixed critical temperature $T_c = 1 / (a_c\!\cdot\!N_t)$ implies a smaller lattice spacing, which in turn corresponds to the weaker bare coupling (larger $\be_F^{(c)}$) shown in \fig{fig:Wpoly_Nt}.

These smaller lattice spacings are known~\cite{Luscher:2009eq} to reduce the efficiency with which the HMC algorithm samples topological sectors characterized by an integer topological charge $Q$.
The small differences visible in \fig{fig:Wpoly_Nt} between the high- and low-start $N_t = 12$ results, for five $16.2 \leq \be_F \leq 17$ on the weak-coupling side of the transition, are related to this topological freezing: these high- and low-start streams are frozen in different sectors with $Q = 0$ and $Q = -1$, respectively. 
While we observe better topological sampling at the $N_t = 12$ transition $\be_F^{(c)} \approx 16$, we will need to monitor this behavior carefully when studying the $N_t \to \infty$ continuum limit of the transition in future work.
After accounting for this topological effect, there is no sign of hysteresis in \fig{fig:Wpoly_Nt}, as we discuss further in \secref{sec:pg_order}.

\begin{figure}[tbp]
  \includegraphics[width=\linewidth]{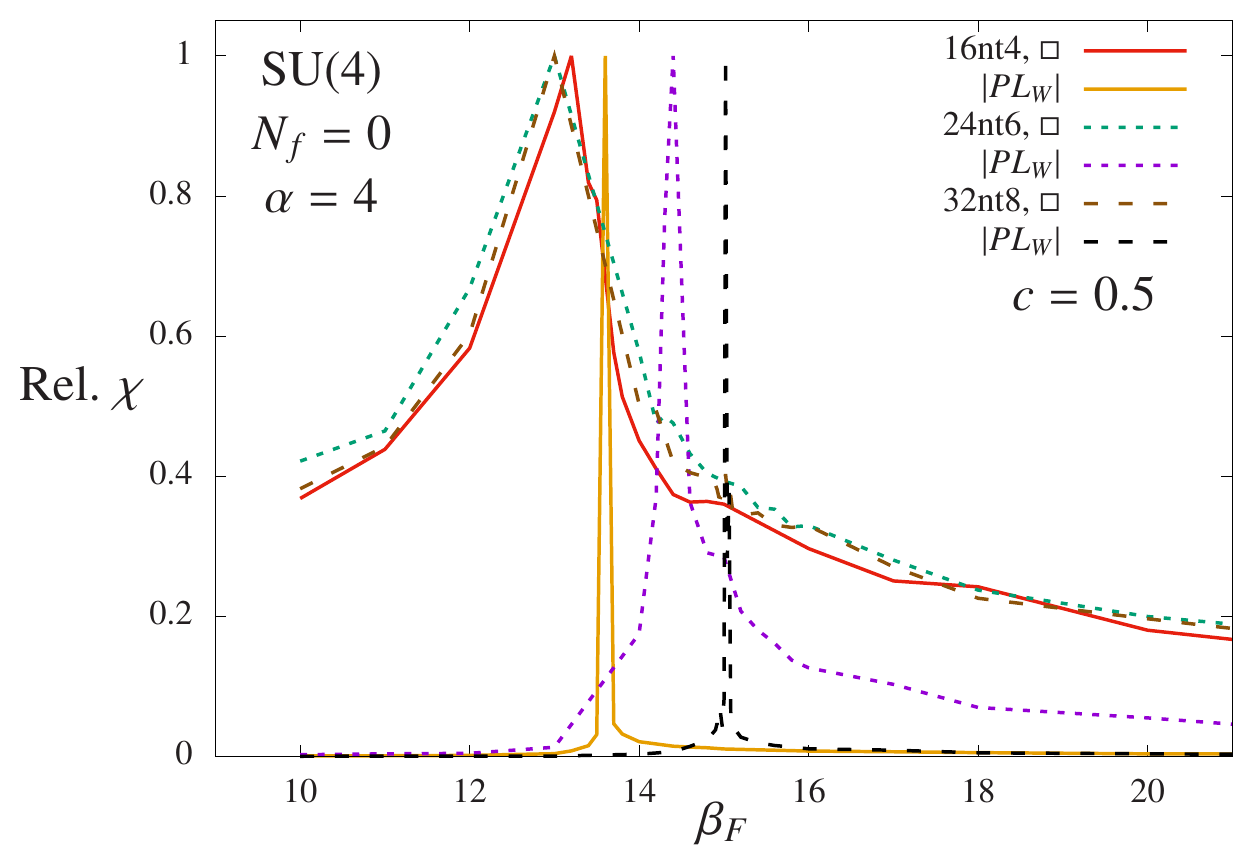}
  \caption{\label{fig:pg_suscept}Pure-gauge SU(4) plaquette ($\Box$) and Wilson-flowed Polyakov loop ($|PL_W|$) susceptibilities $\chi$ plotted vs.\ the bare gauge coupling $\be_F$.  We superimpose results for lattice volumes $16^3\X 4$ (solid), $24^3\X 6$ (dotted) and $32^3\X 8$ (dashed lines) with aspect ratio $\al \equiv L / N_t = 4$.  For clarity we normalize each data set by its maximum peak height, and draw only lines connecting the omitted data points.  $N_t \geq 6$ is required to clearly separate the bulk transition signalled by $\chi_{\Box}$ from the thermal confinement transition signalled by $\chi_{|PL_W|}$.}
\end{figure}

This larger lattice spacing results in larger discretization artifacts, which only become unmanageable if the coupling becomes sufficiently strong to cause a zero-temperature bulk transition into a lattice phase with no continuum limit.
This zero-temperature transition occurs around the same $\be_F \approx 13$ for all $N_t$, and is signalled by a peak in the plaquette susceptibility $\chi_{\Box}$, as opposed to the peak in the Wilson-flowed Polyakov loop susceptibility $\chi_{|PL_W|}$ that is one signal of the confinement transition.
In \fig{fig:pg_suscept} we compare these two susceptibilities on the same set of axes for lattice volumes $16^3\X 4$, $24^3\X 6$ and $32^3\X 8$, each with aspect ratio $L / N_t = 4$.
Because the height of the peak in $\chi_{|PL_W|}$ is orders of magnitude larger than that in $\chi_{\Box}$, we plot the relative susceptibilities obtained by normalizing each data set by the maximum height of its respective peak.

In \fig{fig:pg_suscept} we can see that the $N_t = 4$ confinement transition at $\be_F^{(c)} \approx 13.6$ is dangerously close to the bulk transition at $\be_F \approx 13.2$.
We will therefore need to be wary of including $N_t = 4$ in $N_t \to \infty$ continuum extrapolations, which was also the case for older studies using $\be_A = 0$~\cite{Wingate:2000bb, Datta:2009jn}.
So although we can expect reduced discretization artifacts thanks to our improved fundamental--adjoint gauge action with negative $\be_A = -\be_F / 4$, this improvement appears insufficient to allow us to rely on smaller, cheaper lattice volumes.
Already for $N_t = 6$ we can see a much healthier separation between the two transitions in \fig{fig:pg_suscept}, which improves as $N_t$ increases thanks to the $N_t$-dependence of the thermal confinement transition in contrast to the $N_t$-independence of the bulk transition.
For our ongoing studies of the latent heat and other properties of the stealth dark matter confinement transition, we therefore plan to carry out continuum extrapolations using $N_t = 6$, 8 and 12.
These continuum extrapolations are not crucial for our present task of determining the dynamical SU(4) phase diagram, so for the remainder of this work we will focus on $N_t = 8$ as the largest temporal extent for which we have already accumulated a great deal of numerical data.

\subsection{\label{sec:pg_order}Order of the transition} 
The final goal of our small-scale pure-gauge calculations is to confirm our prior knowledge that the SU(4) confinement transition seen above is indeed strongly first order rather than continuous.
We do this employing the same HMC algorithm, lattice volumes and statistics that we will use in the dynamical case, in order to illuminate the quality of signals we may expect to see for a first-order transition with heavy dynamical fermions.

Already in \fig{fig:Wpoly_Nt} we saw that the Wilson-flowed Polyakov loop magnitude and the Wilson-flowed $E(t)$ anisotropy do not show any sign of hysteresis for aspect ratio $L / N_t = 2$.
This remains true for larger aspect ratios as well.
While hysteresis in the thermodynamic limit $L \to \infty$ can be expected for a strongly first-order transition, its absence for these lattice volumes does not imply a continuous transition in the infinite-volume continuum theory of interest.

\begin{figure}[tbp]
  \includegraphics[width=\linewidth]{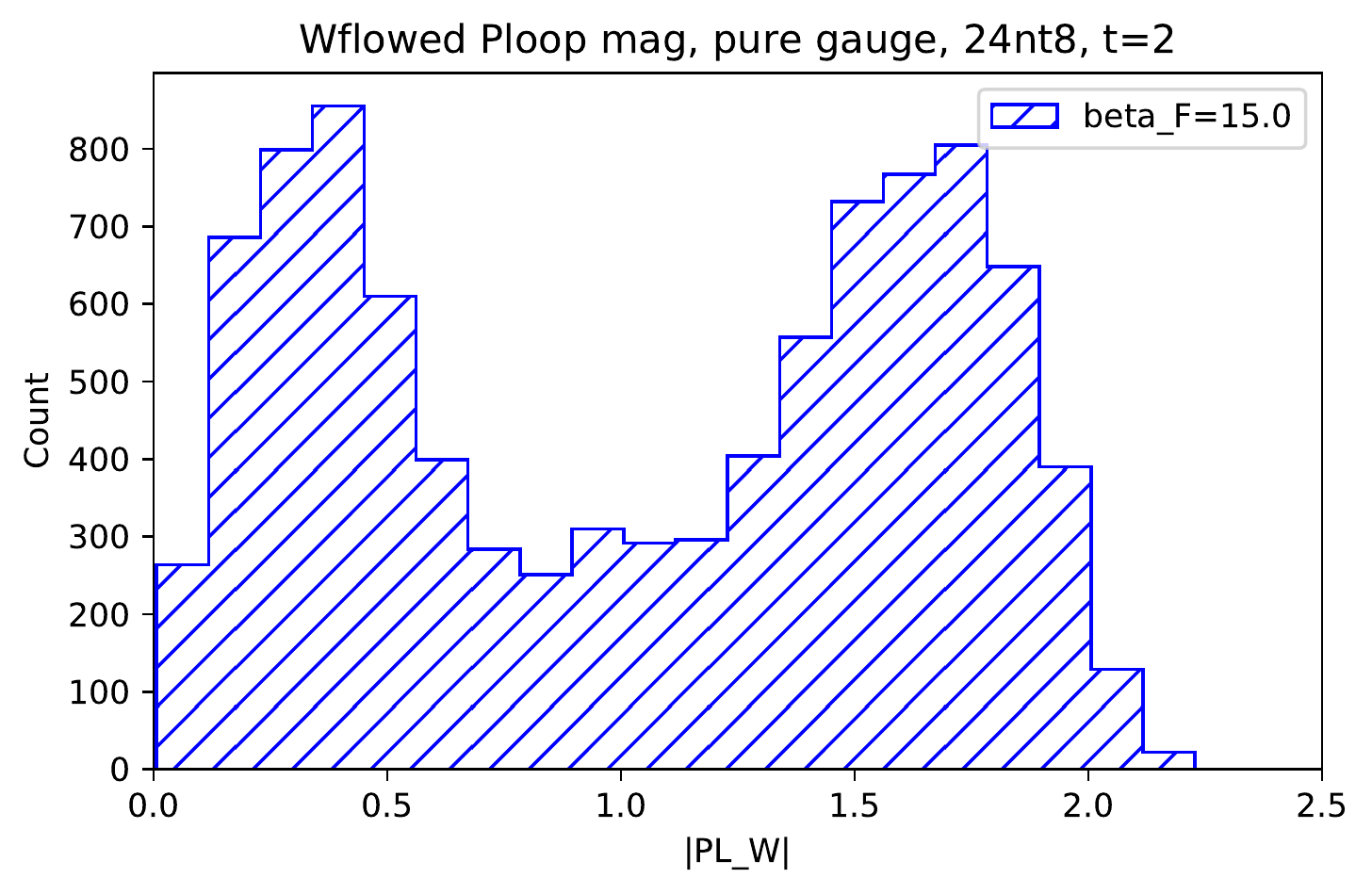}
  \caption{\label{fig:Wpoly_hist-pg}This double-peaked structure in the histogram of Wilson-flowed Polyakov loop magnitude $|PL_W|$ measurements on pure-gauge $24^3\X 8$ lattices with $\be_F = 15.0$ is clear confirmation of a first-order confinement transition.}
\end{figure}

Indeed, from other observables we do have evidence confirming the known first-order nature of the pure-gauge SU(4) confinement transition.
In particular, \fig{fig:Wpoly_hist-pg} shows the histogram of Wilson-flowed Polyakov loop magnitude $|PL_W|$ measurements on $24^3\X 8$ lattices at $\be_F = 15.0$ near the confinement transition.
The histogram features two clearly separated peaks, with approximately the same height, which is characteristic of the confined/deconfined phase coexistence at a first-order transition.
This double-peaked structure is clear confirmation that our calculations suffice to reproduce the known first-order SU(4) confinement transition.

\begin{figure}[tbp]
  \includegraphics[width=\linewidth]{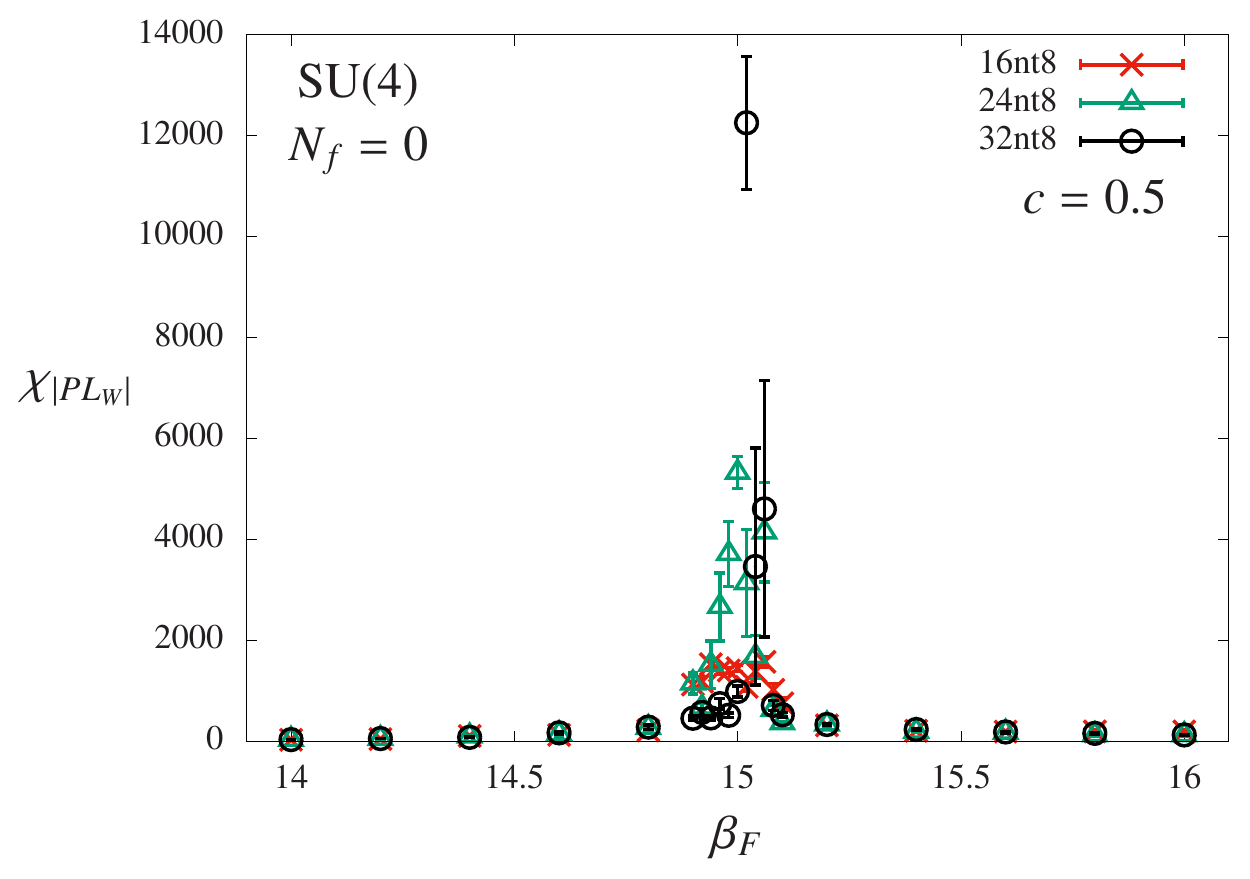}
  \caption{\label{fig:Wpoly_suscept-pg}Peaks in the susceptibility of the Wilson-flowed Polyakov loop magnitude, $\chi_{|PL_W|}$, are consistent with the expected first-order scaling $\chi_{\text{max}} \propto L^3$, for pure-gauge SU(4) lattice ensembles with $N_t = 8$ and aspect ratios $L / N_t = 2$, 3 and 4.}
\end{figure}

A familiar means of determining the order of a confinement transition is to investigate how the maximum height $\chi_{\text{max}}$ of the (Wilson-flowed) Polyakov loop susceptibility peak scales with the spatial lattice volume $L^3$.
A first-order transition is characterized by direct volume scaling $\chi_{\text{max}} \propto L^3$, in contrast to both the critical scaling $\chi_{\text{max}} \propto L^{3b}$ of a second-order transition with critical exponent $b \ne 1$ and the $L$-independence of a continuous crossover~\cite{Imry:1980zz, Fisher:1982xt, Binder:1984llk, Challa:1986sk, Fukugita:1989yb}.
In \fig{fig:Wpoly_suscept-pg} we present the $|PL_W|$ susceptibility peaks for our pure-gauge $N_t = 8$ ensembles with aspect ratios $L / N_t = 2$, 3 and 4, which are consistent with the expected first-order volume scaling.

However, with the lattice volumes and statistics available to us it is difficult to quantitatively verify the volume scaling that would confirm a first-order transition.
In addition to the large uncertainties around the transition,\footnote{Such large uncertainties around first-order transitions are a generic challenge for Markov-chain Monte Carlo calculations, motivating alternate approaches such as density-of-states techniques~\cite{Langfeld:2015fua}.} the peak will occur at slightly different critical $\be_F^{(c)}$ for each different $L$, and the values of $\be_F$ we have sampled may not exactly match these critical couplings.
The situation is similar for the Wilson-flowed Polyakov loop kurtosis [\eq{eq:kurtosis}], which suffers from even larger uncertainties.
Robustly determining these peak locations and heights is usually done through multi-ensemble reweighting~\cite{Kuramashi:2020meg, Ferrenberg:1988yz}, which we have not yet attempted.
Instead, we will rely on our other evidence for a first-order transition, and take \fig{fig:Wpoly_suscept-pg} as an indication of the behavior we should expect to see for a first-order confinement transition in stealth dark matter with dynamical fermions.

\begin{figure}[tbp]
  \includegraphics[width=\linewidth]{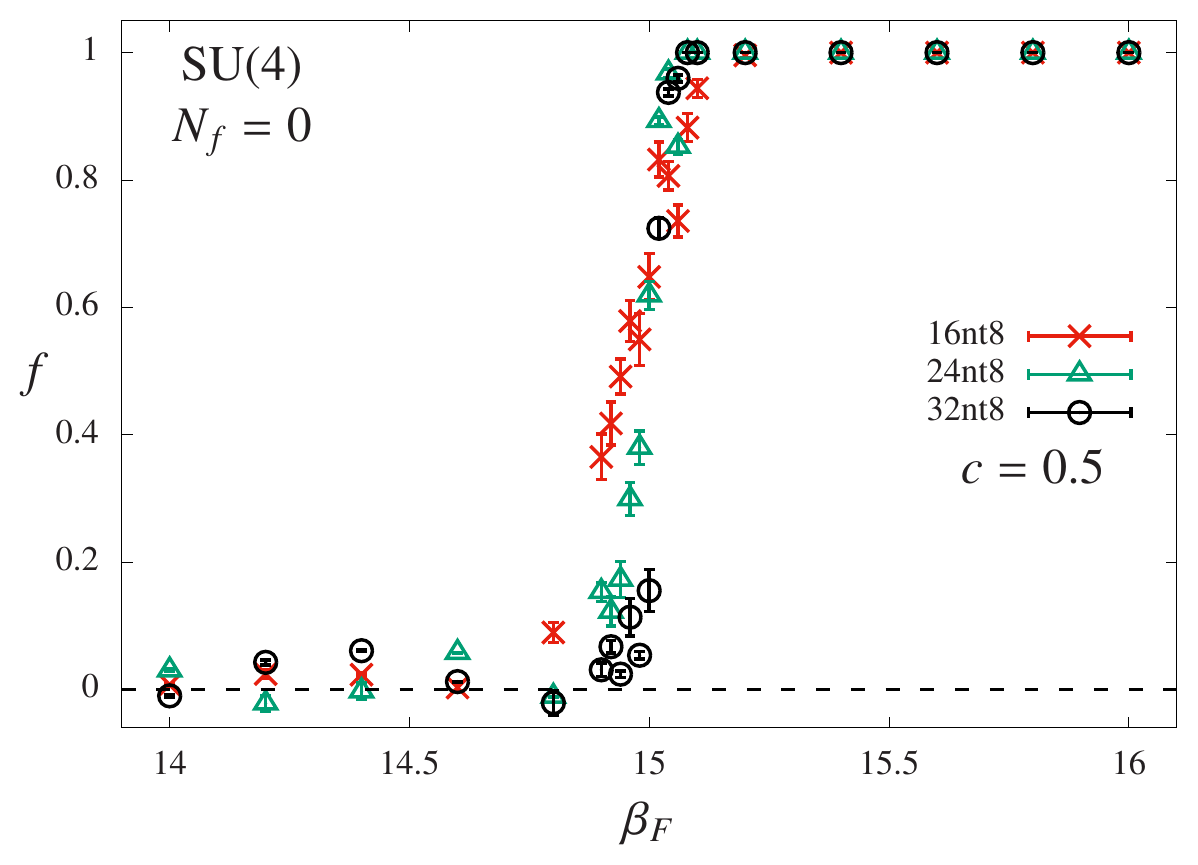}
  \caption{\label{fig:Wdeconf_frac-pg}The deconfinement fraction $f$ from \eq{eq:deconf_frac} with uncertainties obtained as described in the text, for pure-gauge SU(4) lattice ensembles with $N_t = 8$ and aspect ratios $L / N_t = 2$, 3 and 4.  The more rapid change from the $f \to 1$ deconfined limit to the $f \to 0$ confined limit with increasing $L$ is consistent with a discontinuous first-order transition in the $L \to \infty$ thermodynamic limit.}
\end{figure}

To the same end, in \fig{fig:Wdeconf_frac-pg} we show the $L$ dependence of the deconfinement fraction $f$ for the same $N_t = 8$ ensembles with aspect ratios $L / N_t = 2$, 3 and 4.
In \eq{eq:deconf_frac} we normalized the deconfinement fraction so that $f \to 1$ in the deconfined phase and $f \to 0$ in the confined phase.
These limits are clearly seen in \fig{fig:Wdeconf_frac-pg}, up to some residual fluctuations around zero in the $\be_F < \be_F^{(c)}$ confined regime.
The key feature consistent with the first-order nature of the pure-gauge SU(4) confinement transition is that the change between these two limits becomes more rapid as $L$ increases, eventually becoming discontinuous in the $L \to \infty$ thermodynamic limit.
This is another feature of a first-order transition that we will monitor in the case of the stealth dark matter confinement transition, to which we now turn.

\section{\label{sec:4f}Dynamical $N_f = 4$ mass dependence} 
We now consider the more challenging task of studying stealth dark matter by coupling SU(4) lattice gauge theory to $N_f = 4$ degenerate dynamical fermions.
Compared to pure-gauge SU($N$) theories, much less work has been done to investigate finite-temperature dynamics with $N > 3$ and dynamical fermions. 
Reference~\cite{Ayyar:2018ppa} investigates $N_f = 2$ for $3 \leq N \leq 5$ to explore the approach to the large-$N$ limit, while \refcite{DeGrand:2018tzn} also considers $N_f = 2$ for SU(4), as a limit of a theory with multiple fermion representations motivated by a composite Higgs model with partial compositeness.

Compared to composite Higgs studies in which some of the fermions must be massless and others are generically light in order to produce near-conformal dynamics, our task is simplified by considering relatively heavy fermions corresponding to the upper-right corner of the `Columbia plot' in \fig{fig:columbia}.
As described in \secref{sec:lattice}, we consider $\am = \left\{0.05, 0.1, 0.2, 0.4\right\}$, with the smallest $\am = 0.05$ chosen to overlap with the masses considered by previous lattice studies of stealth dark matter~\cite{Appelquist:2015yfa, Appelquist:2015zfa}.
The largest $\am = 0.4$ turns out to be the only one for which we observe a first-order confinement transition.
After presenting our results for the mass dependence of the transition, we will convert these values of \am into ratio of dark pion and dark vector meson masses, $M_P / M_V$, for more direct comparison with Refs.~\cite{Appelquist:2015yfa, Appelquist:2015zfa}.

\subsection{$N_t = 8$ transition results} 
\begin{figure}[tbp]
  \includegraphics[width=\linewidth]{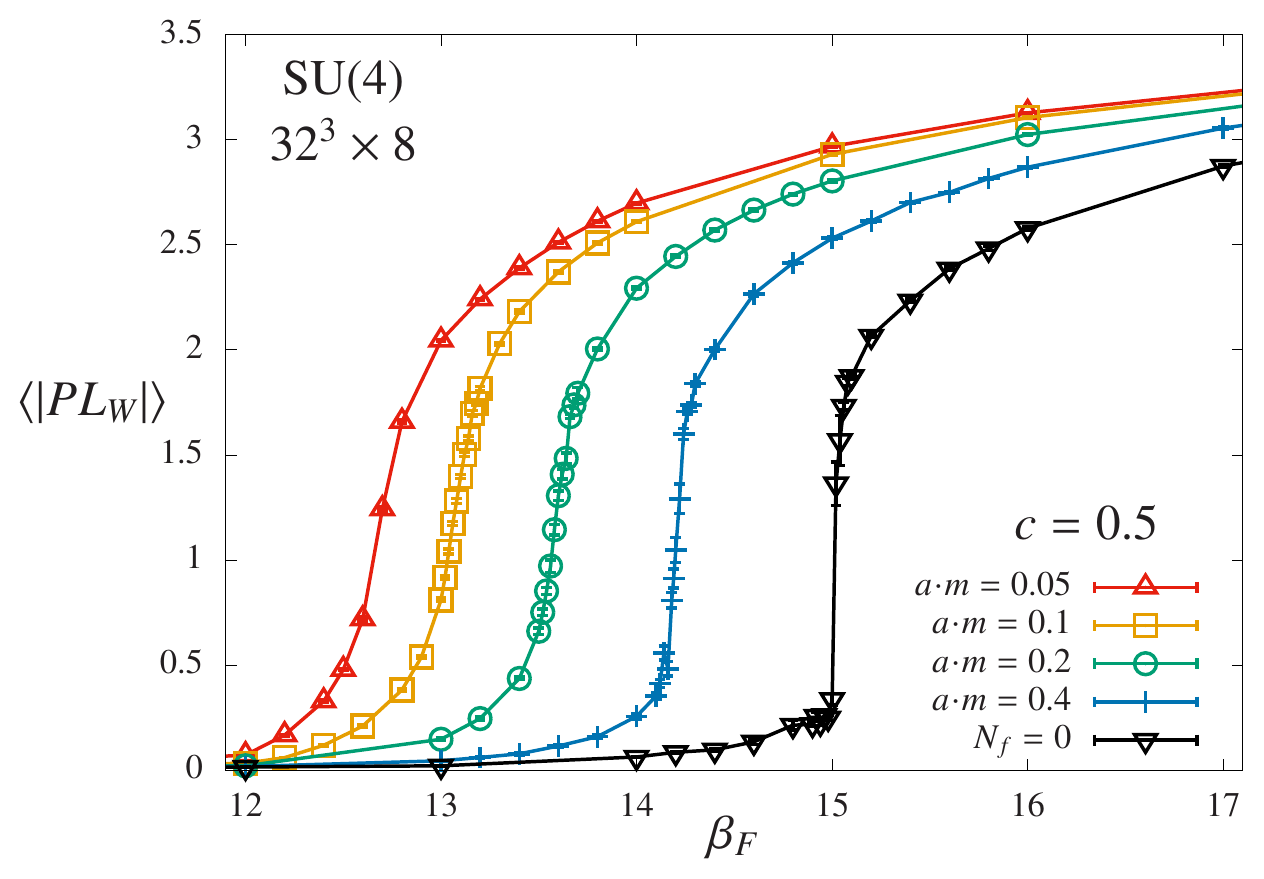}\\[12 pt]
  \includegraphics[width=\linewidth]{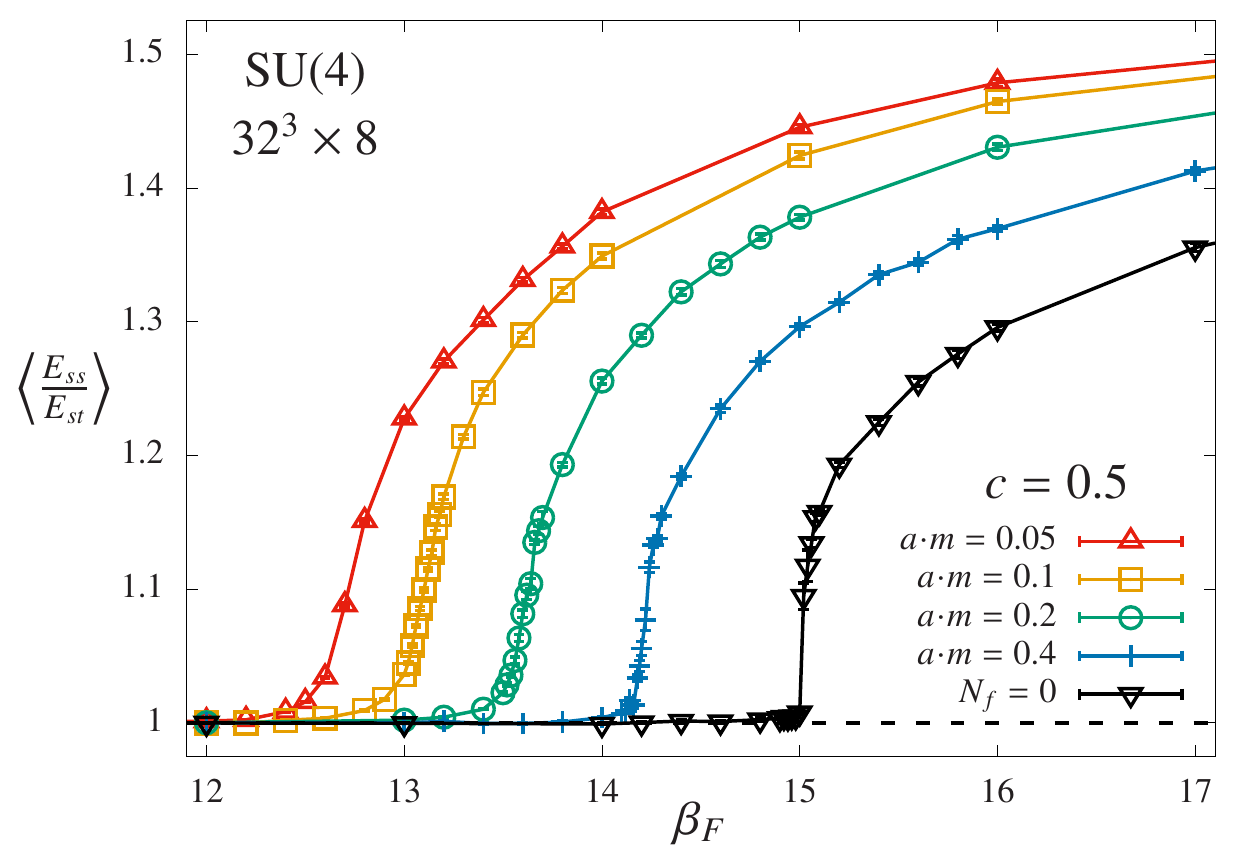}
  \caption{\label{fig:Wpoly_mass}Dependence of the four-flavor SU(4) (pseudo-)critical coupling $\be_F^{(c)}$ on the bare fermion mass $\am$, including the $\am \to \infty$ limit corresponding to the `$N_f = 0$' pure-gauge theory.  As \am decreases, confinement occurs at steadily stronger couplings (smaller $\be_F^{(c)}$), as shown by both the Wilson-flowed Polyakov loop magnitude $|PL_W|$ (top) and the Wilson-flowed $E(t)$ anisotropy (bottom).  Here we show only results for lattice volume $32^3\X 8$, with lines connecting points to guide the eye.}
\end{figure}

As for the pure-gauge limit in \secref{sec:0f}, we begin by briefly considering the critical coupling $\be_F^{(c)}$ of the thermal confinement transition of stealth dark matter.
Since the dependence on the temporal extent of the lattice $N_t$ is similar in both cases, in \fig{fig:Wpoly_mass} we focus on the bare fermion mass $\am$ dependence of $\be_F^{(c)}$ for $N_t = 8$, including the $\am \to \infty$ pure-gauge limit.
As expected, lighter dynamical fermions more effectively screen the gauge interactions, requiring stronger bare couplings (smaller $\be_F$) to produce the transition.
Figure~\ref{fig:Wpoly_mass} shows this for both the Wilson-flowed Polyakov loop magnitude $|PL_W|$ and the Wilson-flowed $E(t)$ anisotropy.
From both these results and the corresponding $|PL_W|$ susceptibility peaks discussed below we can easily read off $\be_F^{(c)} \approx \left\{12.7, 13.1, 13.6, 14.2\right\}$ for $\am = \left\{0.05, 0.1, 0.2, 0.4\right\}$.
Notably, even though $\am = 0.4$ is rather heavy, dynamical fermions with this mass still produce a significant shift in the critical coupling for confinement, compared to the pure-gauge $\be_F^{(c)} \approx 15.0$.
While this shift can be predicted by a simple hopping parameter expansion~\cite{Hasenfratz:1993az}, it indicates that the fermions are not so heavy as to be effectively quenched.

\begin{figure}[tbp]
  \includegraphics[width=\linewidth]{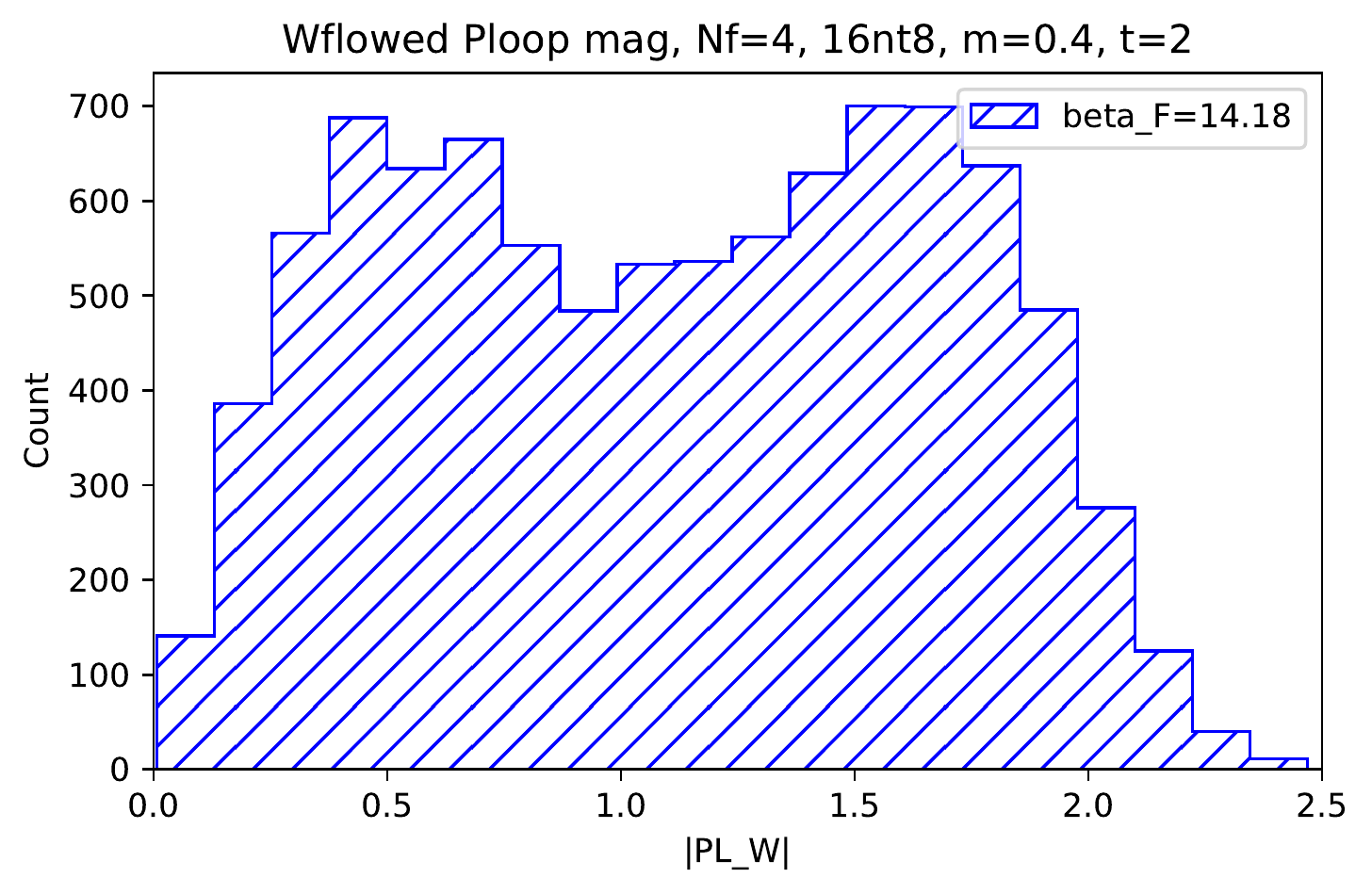}
  \caption{\label{fig:Wpoly_hist-4f}A double-peaked structure in the histogram of Wilson-flowed Polyakov loop magnitude $|PL_W|$ measurements on dynamical $\am = 0.4$ lattices with volume $16^3\X 8$ and $\be_F = 14.18$, evidence for a first-order phase transition at this mass.}
\end{figure}

Since we observed no hysteresis for these quantities in the pure-gauge case in \fig{fig:Wpoly_Nt}, it is not surprising that none of our dynamical $N_f = 4$ streams exhibit any hysteresis, either.
For this reason we have simplified \fig{fig:Wpoly_mass} by including only high-start results.
An initial sign of a first-order transition for $\am = 0.4$ comes from \fig{fig:Wpoly_hist-4f}, which shows a double-peaked structure consistent with confined/deconfined phase coexistence at a first-order transition.
Compared to the pure-gauge histogram in \fig{fig:Wpoly_hist-pg}, the valley between the two peaks is much less dramatic in this dynamical case, and we see no two-peak structure for any of our $\am \leq 0.2$ ensembles.
This suggests that $\am = 0.2$ is sufficiently small to move the system out of the heavy-mass first-order region that appears to contain $\am = 0.4$.

\begin{figure*}[tbp]
  \includegraphics[width=0.45\linewidth]{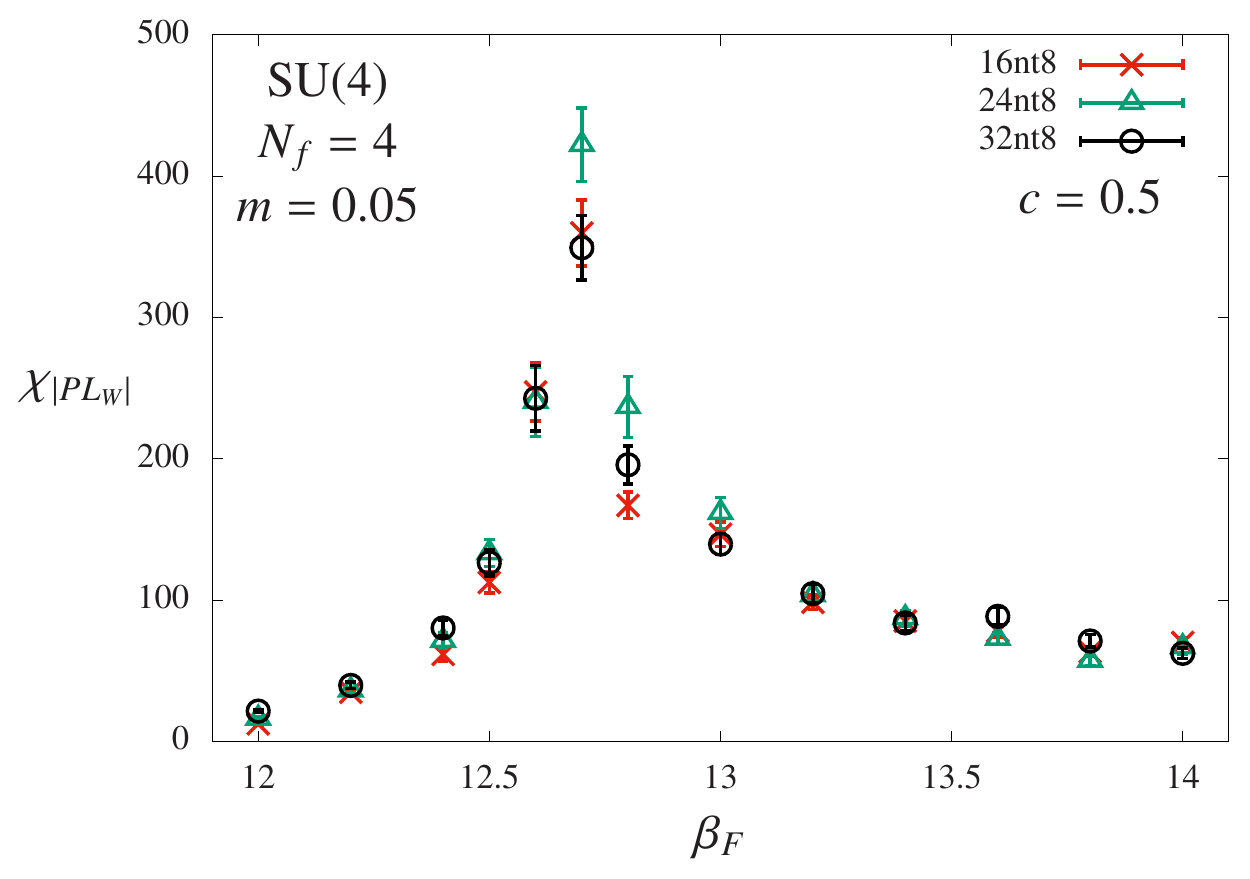}\hfill \includegraphics[width=0.45\linewidth]{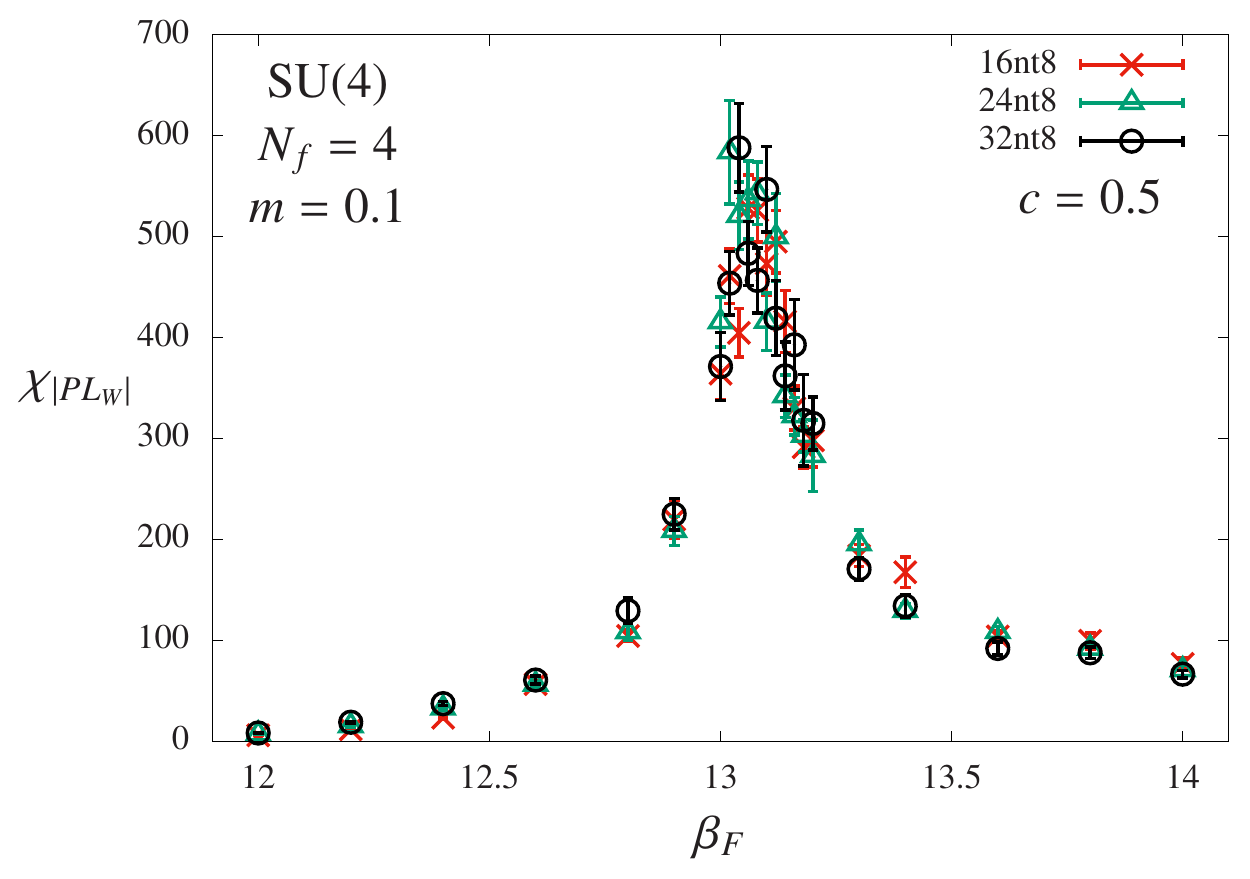} \\[12 pt]
  \includegraphics[width=0.45\linewidth]{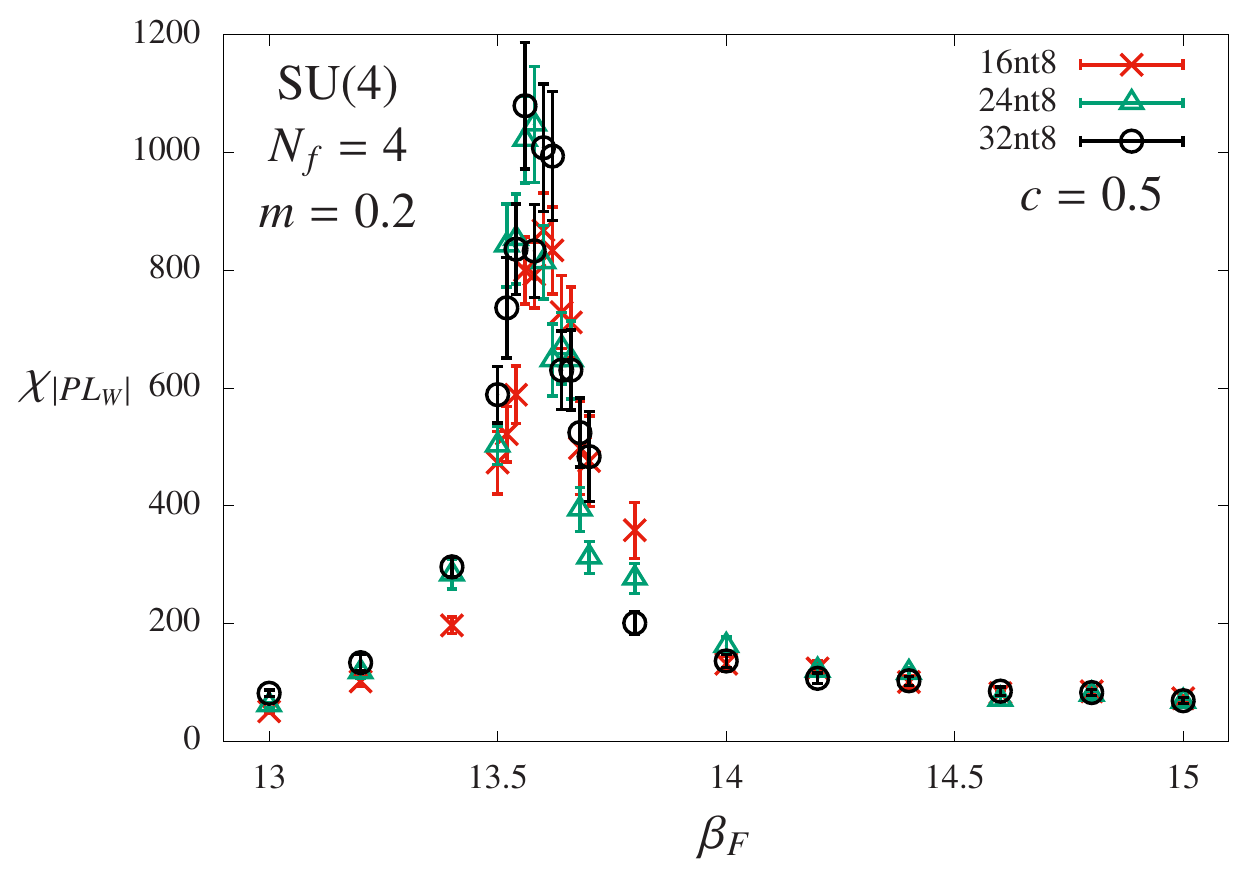}\hfill  \includegraphics[width=0.45\linewidth]{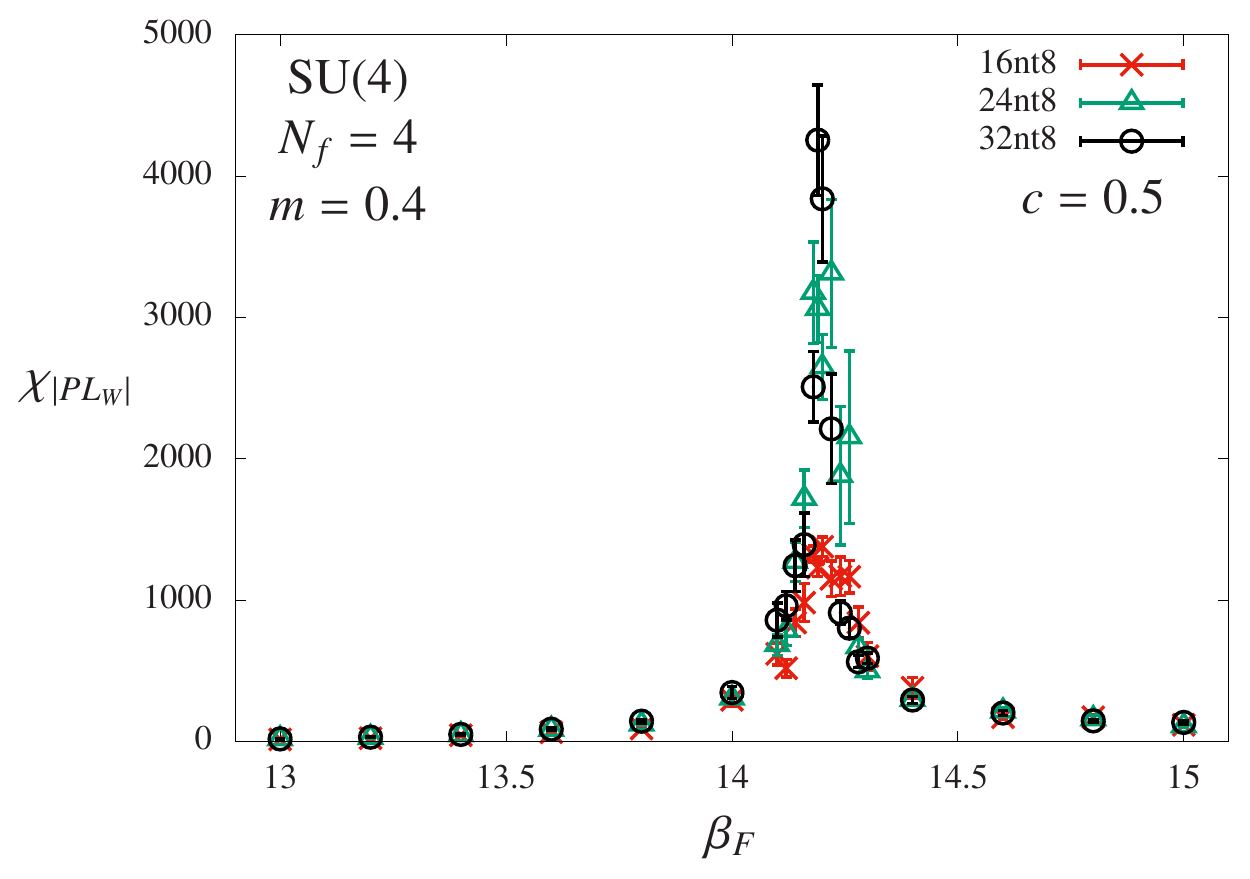}
  \caption{\label{fig:Wpoly_suscept-4f}Peaks in the susceptibility of the Wilson-flowed Polyakov loop magnitude, $\chi_{|PL_W|}$, for SU(4) lattice ensembles with $N_t = 8$, $L / N_t = 2$, 3 and 4, and dynamical fermion masses $\am = 0.05$ (upper left), 0.1 (upper right), 0.2 (lower left) and 0.4 (lower right).  Only the $\am = 0.4$ results could be consistent with the maximum peak heights exhibiting the volume scaling $\chi_{\text{max}} \propto L^3$ of a first-order transition.  The range of the vertical axes depends strongly on $\am$, while the horizontal axes always span $\De \be_F = 2$.}
\end{figure*}
\begin{figure*}[tbp]
  \includegraphics[width=0.45\linewidth]{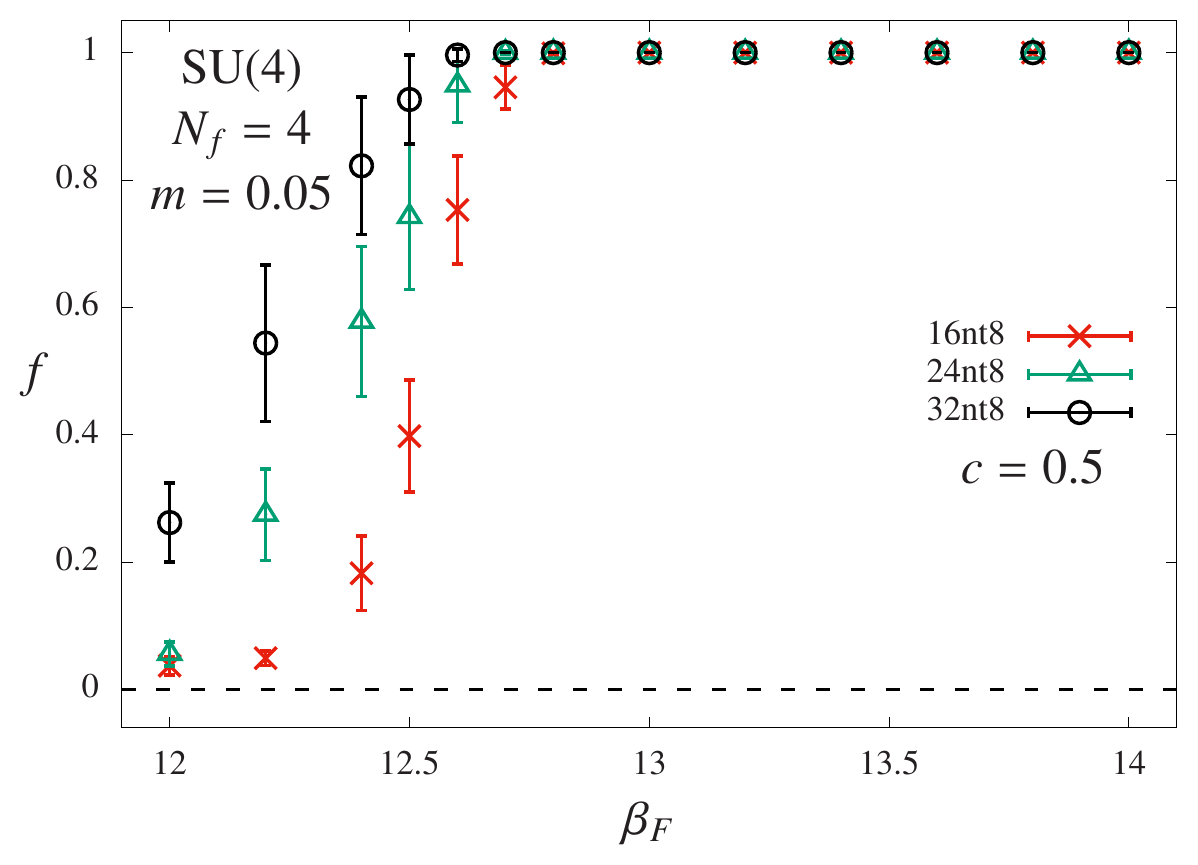}\hfill \includegraphics[width=0.45\linewidth]{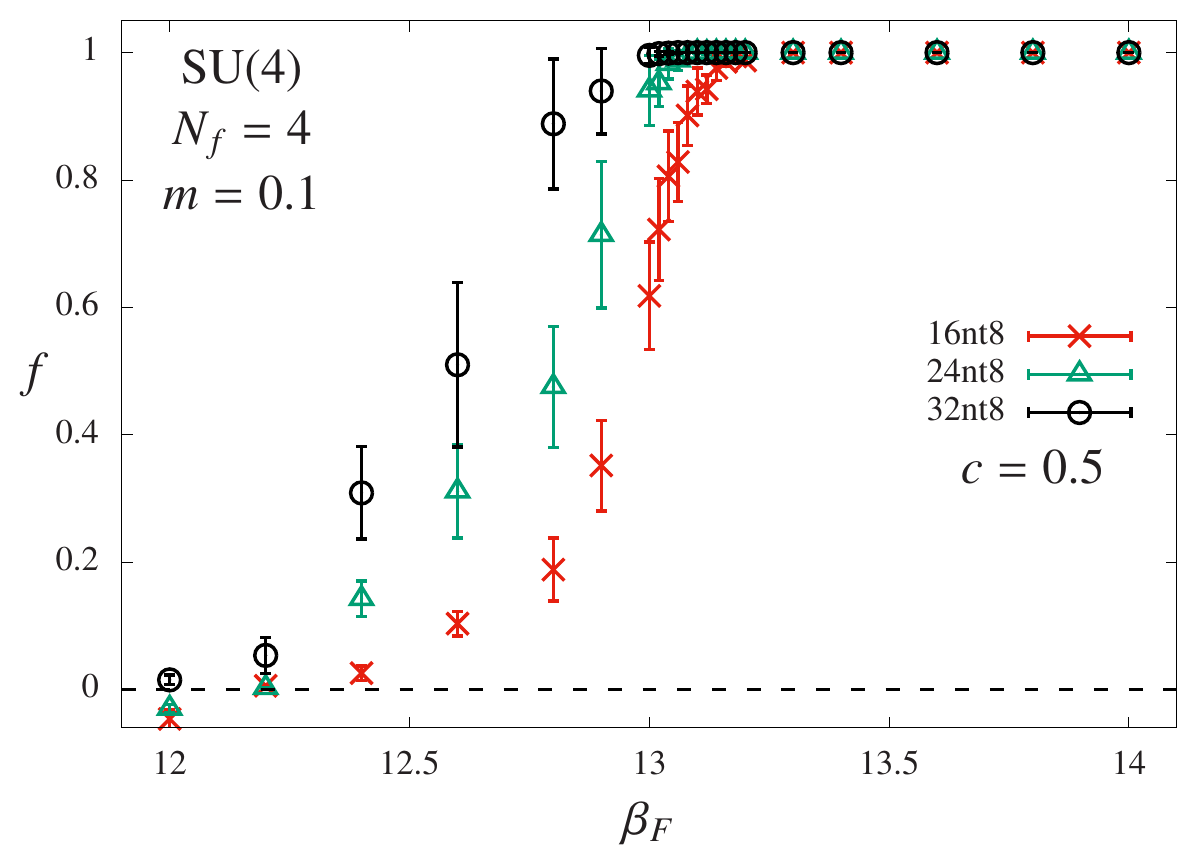} \\[12 pt]
  \includegraphics[width=0.45\linewidth]{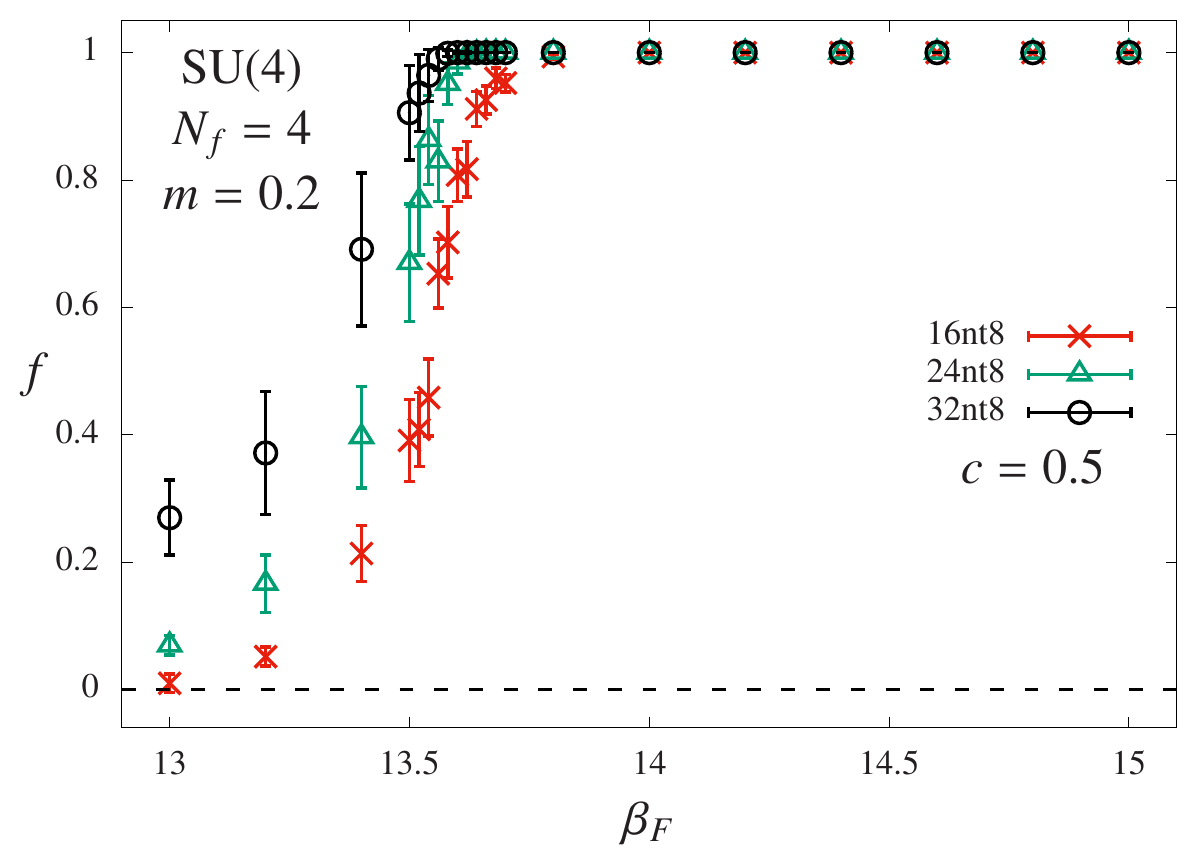}\hfill  \includegraphics[width=0.45\linewidth]{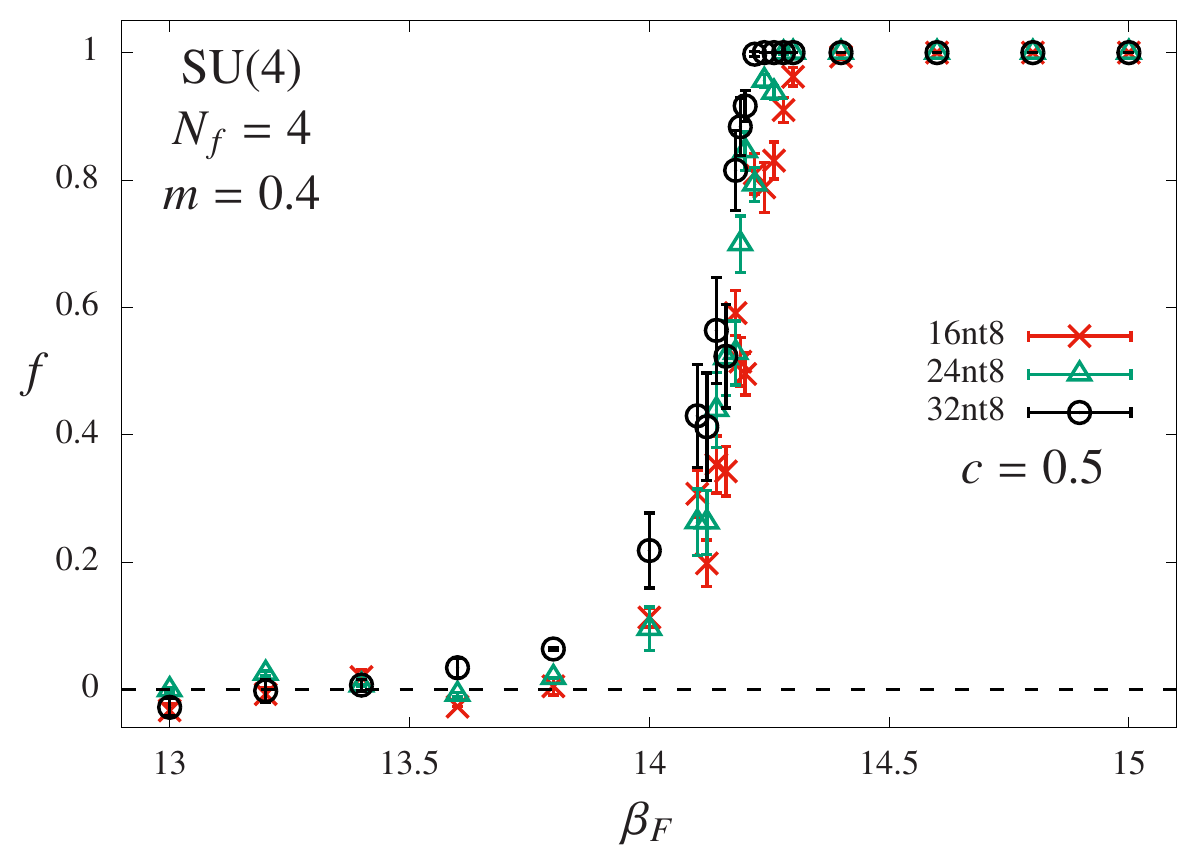}
  \caption{\label{fig:Wdeconf_frac-4f}The deconfinement fraction $f$ from \eq{eq:deconf_frac} with uncertainties obtained as described in the text, for SU(4) lattice ensembles with $N_t = 8$, $L / N_t = 2$, 3 and 4, and dynamical fermion masses $\am = 0.05$ (upper left), 0.1 (upper right), 0.2 (lower left) and 0.4 (lower right).  Only the $\am = 0.4$ results could be consistent with a first-order transition in the $L \to \infty$ thermodynamic limit.}
\end{figure*}

In Figs.~\ref{fig:Wpoly_suscept-4f} and \ref{fig:Wdeconf_frac-4f} we more comprehensively compare our four dynamical masses $\am = \left\{0.05, 0.1, 0.2, 0.4\right\}$, considering the same $L$ dependence of the Wilson-flowed Polyakov loop susceptibility $\chi_{|PL_W|}$ and deconfinement fraction $f$ as shown for the pure-gauge theory in Figs.~\ref{fig:Wpoly_suscept-pg} and \ref{fig:Wdeconf_frac-pg}, respectively.
We again focus on $N_t = 8$ with aspect ratios $L / N_t = 2$, 3 and 4, generating a higher density of ensembles around the transition for each case, except $\am = 0.05$ which is clearly a smooth crossover.

For the susceptibility $\chi_{|PL_W|}$ in \fig{fig:Wpoly_suscept-4f}, the height of the $32^3\X 8$ peaks increases by an order of magnitude as the mass increases from $\am = 0.05$ to $0.4$, though that last case still remains significantly below the scale of the pure-gauge peak in \fig{fig:Wpoly_suscept-pg} (again indicating that the fermions are not so heavy as to be effectively quenched).
In combination with the fixed width of the horizontal axes, the increasing range of the vertical axes produces narrower-looking peaks as \am increases.
As discussed in \secref{sec:obs}, the Polyakov loop is no longer a true order parameter in the presence of dynamical fermions in the fundamental representation, and it may not remain a useful observable if \am is made too small.
Figures~\ref{fig:Wpoly_mass} and \ref{fig:Wpoly_suscept-4f} empirically show that $|PL_W|$ remains a good indicator of the phase structure throughout the range of relatively large \am we consider.

\begin{figure}[tbp]
  \includegraphics[width=\linewidth]{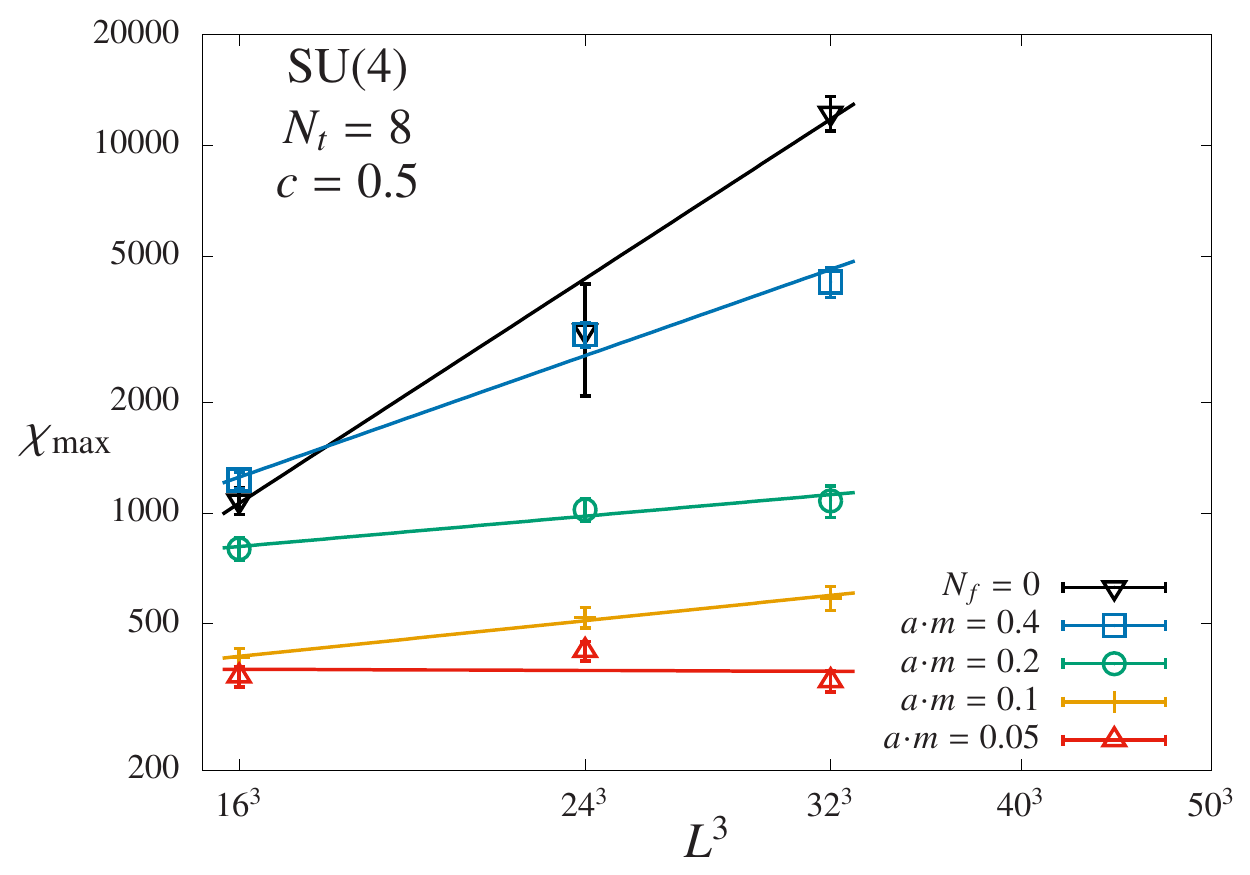}
  \caption{\label{fig:Wpoly_suscept_scale}Wilson-flowed Polyakov loop susceptibility peak heights, $\chi_{\text{max}}$, plotted against the spatial volume, $L^3$, on log--log axes.  The straight lines are power-law fits, which for $N_f = 0$ and $\am = 0.4$ are roughly consistent with the volume scaling $\chi_{\text{max}} \propto L^3$ of a first-order transition.}
\end{figure}

The key feature in \fig{fig:Wpoly_suscept-4f} is the $L$ dependence of the maximum peak heights $\chi_{\text{max}}$, which as discussed in \secref{sec:pg_order} is difficult to determine given the increasing uncertainties around the transition and the non-zero $\De \be_F = 0.02$ separating ensembles in the transition region.
In \fig{fig:Wpoly_suscept_scale} we plot $\chi_{\text{max}}$ against the spatial volume $L^3$, on log--log axes, and overlay power-law fits, $\chi_{\text{max}} \propto L^{3b}$.
Since we have not yet attempted the multi-ensemble reweighting~\cite{Kuramashi:2020meg, Ferrenberg:1988yz} that is likely necessary to reliably estimate the true heights of the susceptibility peaks, the exponents $b$ are not well determined.
Given these limitations, the values $b \sim 1.2$ and $b \sim 0.6$ that we obtain for the pure-gauge theory and $\am = 0.4$, respectively, are consistent with the volume scaling $\chi_{\text{max}} \propto L^3$ of a first-order transition.
For the lighter $\am = \{0.2, 0.1, 0.05\}$ we find much smaller $b \sim \{0.2, 0.2, 0.0\}$, respectively, consistent with the $L$ independence expected for a continuous crossover.

Figure~\ref{fig:Wdeconf_frac-4f} provides another consistency check supporting the same conclusion.
The most notable feature of these plots is the very slow decrease of the $32^3\X 8$ deconfinement fraction for $\am \leq 0.2$ on the $\be_F < \be_F^{(c)}$ confined side of the critical coupling indicated by the $|PL_W|$ susceptibility peaks.
Empirically, for $\am \leq 0.2$ we also observe $f \approx 1$ for the $32^3\X 8$ ensembles that produce the largest susceptibilities $\chi_{|PL_W|}$, while $\am = 0.4$ and the pure-gauge theory both produce values [$f = 0.906(41)$ and $f = 0.795(14)$, respectively] significantly below unity. 
Again, the $\am = 0.4$ results are the only ones qualitatively consistent with the pure-gauge behavior in \fig{fig:Wdeconf_frac-pg}.
While the development of a discontinuity in the $L \to \infty$ thermodynamic limit is not obvious in this case, the clear contrast with the $\am \leq 0.2$ results still suggests a change to a first-order transition for $\am = 0.4$.

\subsection{Zero-temperature spectroscopy} 
Our final task in this work is to parameterize the \am discussed above in a convenient form for comparison with previous lattice studies of stealth dark matter~\cite{Appelquist:2015yfa, Appelquist:2015zfa}.
We do this by computing the ratio of dark pion and dark vector meson masses, $M_P / M_V$, which requires `zero-temperature' lattice calculations with $N_t > L$.
We carry out these zero-temperature calculations at the $\be_F^{(c)}$ of the $N_t = 8$ transitions discussed above, for each bare fermion mass $\am = \left\{0.05, 0.1, 0.2, 0.4\right\}$.
As $N_t$ increases and the corresponding lattice spacing $a_c \simeq 1 / (T_c\!\cdot\!N_t)$ decreases, we will need to consider correspondingly smaller bare masses $\am$ in order to take the $a \to 0$ continuum limit along a `line of constant physics' with fixed $M_P / M_V$.
In this work we restrict ourselves to determining the $M_P / M_V$ corresponding to the $N_t = 8$ transitions.

To determine $M_P$ and $M_V$ we carry out correlated fits of the corresponding two-point staggered correlation functions, over appropriate fit ranges $[t_{\text{min}}, t_{\text{max}}]$. 
We do not include any excited states in our fits, instead considering relatively large $t_{\text{min}}$ to reduce any possible excited-state contamination.
For $\am = 0.05$ and 0.1, we fix $t_{\text{max}} = N_t / 2 = 24$ and combine results for all $t_{\text{min}}$ in the range $10 < t_{\text{min}} < 18$.
For the larger masses $\am = 0.2$ and 0.4, the exponential decay of the correlation functions $C(t) \sim e^{-Mt}$ at large times $t$ can cause the signal in the vector channel to be overwhelmed by statistical noise for $t < N_t / 2$.
This requires that we set a smaller $t_{\text{max}} = 16$, which in turn demands a smaller range of $6 < t_{\text{min}} < 12$.

\begin{table*}[tbp]
  \centering
  \renewcommand\arraystretch{1.2}  
  \addtolength{\tabcolsep}{3 pt}   
  \begin{tabular}{lcc|lll|rc}
    \hline
    \am  & $\be_F$ & Bins & $\sqrt{8t_0} / a$ & $a\!\cdot\!M_P$ & $a\!\cdot\!M_V$ & $M_P / T_c$ & $M_P / M_V$ \\
    \hline
         & 12.4    & 70   & 2.86514(82)       & 0.493225(91)    & 0.7951(85)      &  3.95       & 0.620       \\
    0.05 & 12.6    & 65   & 3.3041(13)        & 0.46419(12)     & 0.7161(30)      &  3.71       & 0.648       \\
         & 12.8    & 60   & 3.7587(16)        & 0.43880(14)     & 0.6443(20)      &  3.51       & 0.681       \\
    \hline
         & 12.8    & 80   & 3.3830(11)        & 0.65305(10)     & 0.8461(14)      &  5.22       & 0.772       \\
    0.1  & 13.0    & 80   & 3.8124(16)        & 0.62368(14)     & 0.78404(77)     &  4.99       & 0.795       \\
         & 13.2    & 80   & 4.2548(20)        & 0.59736(13)     & 0.73213(60)     &  4.78       & 0.816       \\
    \hline
         & 13.4    & 70   & 4.0836(18)        & 0.88465(12)     & 0.98592(24)     &  7.08       & 0.897       \\
    0.2  & 13.6    & 70   & 4.5153(30)        & 0.85889(15)     & 0.94163(29)     &  6.87       & 0.912       \\
         & 13.8    & 78   & 4.9623(38)        & 0.83186(15)     & 0.90249(27)     &  6.65       & 0.922       \\
    \hline
         & 14.0    & 80   & 4.6153(27)        & 1.28724(10)     & 1.34138(17)     & 10.30       & 0.960       \\
    0.4  & 14.2    & 80   & 5.0413(38)        & 1.26126(11)     & 1.31148(17)     & 10.09       & 0.962       \\
         & 14.4    & 70   & 5.5108(47)        & 1.24108(13)     & 1.27758(18)     &  9.93       & 0.971       \\
    \hline
  \end{tabular}
  \caption{\label{tab:spectrum}Results for the Wilson flow scale, pseudoscalar meson mass and vector meson mass for each of our zero-temperature $24^3\X 48$ ensembles, using the stated number of 100-MDTU (10-measurement) bins and also comparing $M_P$ to the $N_t = 8$ critical temperature.  The uncertainties on the individual masses come from correlated fits described in the text.  Rather than propagate these to the ratio $M_P / M_V$, we take the uncertainty on the ratio to be dominated by varying the coupling $\be_F^{(c)} \pm 0.2$ around its critical value for each fermion mass $\am$.  This produces $M_P / M_V = \left\{0.65(3), 0.79(2), 0.91(1), 0.96(1)\right\}$ for $\am = \left\{0.05, 0.1, 0.2, 0.4\right\}$, respectively.}
\end{table*}

Our results for $a\!\cdot\!M_P$ and $a\!\cdot\!M_V$ are compiled in \tab{tab:spectrum}, where for reference we also include results for the scale $\sqrt{8t_0}$ introduced in \refcite{Luscher:2010iy} and defined through the Wilson flow discussed in \secref{sec:obs}.
Following Refs.~\cite{Ce:2016awn, DeGrand:2017gbi, DeGrand:2018tzn}, we define this scale through the condition $\left\{t^2\vev{E(t)}\right\}_{t = t_0} = 0.4$, where the energy density $E(t)$ is evaluated after flow time $t$ using the standard clover construction mentioned in \secref{sec:obs}.
This choice incorporates the leading-order scaling $t^2\vev{E(t)} \sim N$ to generalize the canonical SU(3) value of 0.3 to our SU(4) theory.
For convenience we also record the ratio of the pseudoscalar meson mass to the $N_t = 8$ critical temperature, $M_P / T_c = a\!\cdot\!M_P N_t$.

The results shown in \tab{tab:spectrum} do not include systematic uncertainties related to the choice of fit ranges and possible excited-state contamination or finite-volume effects.
Based on our expectation that the overall uncertainty in the $M_P / M_V$ ratio of interest will be dominated by its dependence on the coupling $\be_F$, we simply set that overall uncertainty by varying $\be_F^{(c)} \pm 0.2$ around the $N_t = 8$ critical value for each fermion mass $\am$.
From the table we can therefore read off $M_P / M_V = \left\{0.65(3), 0.79(2), 0.91(1), 0.96(1)\right\}$ for $\am = \left\{0.05, 0.1, 0.2, 0.4\right\}$, respectively.
We can also see that larger $\be_F$ (smaller lattice spacings) produce larger $M_P / M_V$, confirming that smaller \am will be needed to stay on a line of constant physics when taking the $N_t \to \infty$ continuum limit in future work.

Previous lattice studies of stealth dark matter~\cite{Appelquist:2015yfa, Appelquist:2015zfa} considered the mass range $0.55 \lsim M_P / M_V \lsim 0.77$, using valence Wilson fermions on quenched gauge field configurations.
For the $N_t = 8$ transition, our spectrum results for $\am = 0.1$ lie just above this range, which was our motivation for investigating the $\am = 0.05$ case with $M_P / M_V = 0.65(3)$.
In the bigger picture, we see that the $M_P / M_V > 0.9$ required for stealth dark matter to produce a first-order transition in the early universe is significantly larger than the masses previously considered.
This may have non-trivial implications for the phenomenology of the theory, which we will discuss below and could be explored in future research.

\section{\label{sec:conc}Conclusions and next steps} 
We have presented non-perturbative lattice investigations of the finite-temperature confinement transition of SU(4) stealth dark matter, motivated by the possibility that this early-universe phase transition could have produced a stochastic background of gravitational waves that may be constrained or discovered by future searches.
A first-order transition is required to produce such a stochastic background of gravitational waves, so we have focused on determining the region of parameter space for which the stealth dark matter confinement transition is first order, considering relatively heavy dynamical fermions corresponding to the upper-right corner of the Columbia plot (\fig{fig:columbia}).
The infinite-mass limit reduces to pure-gauge SU(4) Yang--Mills theory, which is known to exhibit a strongly first-order confinement transition~\cite{Lucini:2005vg, Datta:2009jn, Lucini:2012gg}.
We analyzed both the pure-gauge theory and a range of dynamical-fermion masses $0.05 \leq \am \leq 0.4$, finding that heavy masses corresponding to a dark meson mass ratio $M_P / M_V > 0.9$ are required to produce a first-order stealth dark matter confinement transition.

Focusing on finite-temperature transitions for temporal lattice extent $N_t = 8$, we identified three signals of a first-order transition for which our $\am = 0.4$ results exhibit the same qualitative behavior as we observe for the known first-order transition in the pure-gauge limit, in contrast to our other calculations with $\am \leq 0.2$.
First, Figs.~\ref{fig:Wpoly_hist-4f} and \ref{fig:Wpoly_hist-pg} show double-peaked structures in the histogram of Wilson-flowed Polyakov loop magnitude $|PL_W|$ measurements, indicating confined/deconfined phase coexistence.
Second, the $\am = 0.4$ case is the only one for which the $|PL_W|$ susceptibility peaks in \fig{fig:Wpoly_suscept-4f} grow with the spatial lattice volume $L^3$, similar to the pure-gauge peak in \fig{fig:Wpoly_suscept-pg} and as required to be consistent with first-order volume scaling $\chi_{\text{max}} \propto L^3$
Finally, the $\am = 0.4$ deconfinement fraction results in \fig{fig:Wdeconf_frac-4f} are the only set that resemble the pure-gauge case in \fig{fig:Wdeconf_frac-pg} and could be consistent with a discontinuity developing in the $L \to \infty$ thermodynamic limit as required for a first-order transition.

Concluding that heavy bare fermion masses $\am > 0.2$ are required in order to obtain a first-order $N_t = 8$ confinement transition, we carried out zero-temperature dark meson spectroscopy calculations to translate this into the constraint $M_P / M_V > 0.9$ for the dimensionless dark meson mass ratio.
We therefore predict that stealth dark matter will produce a stochastic gravitational wave background only for dark fermion masses significantly heavier than those considered by previous lattice studies of stealth dark matter~\cite{Appelquist:2015yfa, Appelquist:2015zfa}, which corresponded to $0.55 \lsim M_P / M_V \lsim 0.77$.
Even in this heavy-mass regime the dynamical fermions play a significant role, as shown by the mass dependence of the critical coupling in \fig{fig:Wpoly_mass} and the height of the $|PL_W|$ susceptibility peaks in \fig{fig:Wpoly_suscept-4f} compared to \fig{fig:Wpoly_suscept-pg}.
However, such dark fermion masses much larger than the confinement scale, as implied by these large $M_P / M_V > 0.9$, may result in stable dark glueballs that contribute to the relic density~\cite{Kribs:2016cew}, potentially requiring reconsideration of the phenomenology and constraints reported by Refs.~\cite{Appelquist:2015yfa, Appelquist:2015zfa}.

Of course, as discussed in \secref{sec:intro}, we are considering stealth dark matter in the $N_f = 4$ limit where all four dark fermions have the same mass.
While only a small splitting between two pairs of degenerate fermions is required by Big Bang nucleosynthesis, such a splitting could in principle be quite large, without running afoul of other constraints.
For such an $N_f = 2 + 2$ theory, the lighter pair of fermions should produce a smaller meson mass ratio $M_P / M_V$ in the first-order transition region, which needs to be kept in mind when applying collider constraints.
In the future it may be interesting to carry out dedicated finite-temperature lattice calculations exploring this more general $N_f = 2 + 2$ setup.
While this could be done by taking the square root of the staggered-fermion lattice action used in this work, switching to domain-wall fermions should also be considered.

Turning back to the $N_f = 4$ case, we can compare our results for $M_P / T_c$ in \tab{tab:spectrum} with the SU(3) endpoint value $M_P / T_c \gsim 10$ reported by \refcite{Ejiri:2019csa} (for $N_f = 2$ and $N_f = 2 + 1$).
Based on our conclusion that the heavy-mass line of $N_t = 8$ first-order transitions turns into a continuous crossover between $0.2 < \am < 0.4$, we predict an SU(4) endpoint value between $7 \lsim M_P / T_c \lsim 10$.
This is not significantly different than the SU(3) value, though only rough comparisons are possible given that different lattice actions are used and continuum extrapolations have not yet been completed in either case.
In particular, \refcite{Ejiri:2019csa} reports significant changes upon moving from $N_t = 4$ to $N_t = 6$, both of which are smaller than the $N_t = 8$ we consider here.\footnote{Following the completion of our work, a new analysis of the SU(3) $N_f = 2$ endpoint $M_P / T_c$~\cite{Cuteri:2020yke} reported significantly larger $M_P / T_c \approx 18$ for $N_t = 6$ with the same action as \refcite{Ejiri:2019csa}.  As $N_t$ increases, \refcite{Cuteri:2020yke} finds smaller $M_P / T_c \approx 16$ and $15$ for $N_t = 8$ and $10$, respectively.}

With $M_P / M_V > 0.5$, the strongest constraint on the stealth dark matter model is $M_V \gsim 2$~TeV coming from $Z' \to \ell^+ \ell^-$ searches~\cite{Kribs:2018ilo}.
By using the result $a\!\cdot\!M_V \approx 1.3$ for $\am = 0.4$ in \tab{tab:spectrum}, we can relate
\begin{equation*}
  T_c = \frac{1}{\aNt} = \frac{1}{(1.3 / M_V)\!\cdot\!8}
\end{equation*}
to translate this constraint into an estimate for the minimum stealth dark matter critical temperature required to produce gravitational waves,
\begin{equation}
  T_c \gsim \frac{2~\mbox{TeV}}{8\!\cdot\!1.3} \approx 0.2~\mbox{TeV}.
\end{equation}
Recalling from \secref{sec:intro} that the dark baryon may have a mass of hundreds of TeV, we can consider a rough upper bound for $M_V \simeq \MDM / 2$ also in the range of hundreds of TeV, which would imply a critical temperature of tens of TeV.
If supercooling effects are mild enough that the transition temperature $T_*$ is not too much lower than this equilibrium critical temperature, then the peak frequency of the gravitational wave spectrum would likely correspond to a range of frequencies well suited to be probed by the LISA observatory~\cite{Schwaller:2015tja, Caprini:2015zlo, Caprini:2019egz} and the proposed future Einstein Telescope~\cite{Punturo:2010zz}.
While the discovery of such stochastic gravitational waves from the early universe would of course be very exciting, even constraints on their spectrum would place novel new bounds on the viable parameter space of stealth dark matter, likely going beyond what may be possible at colliders and direct-detection experiments.

Looking beyond the predictions discussed above, now that we have located a first-order stealth dark matter confinement transition, the next stage of our work will be to study it in more detail in order to more robustly predict the spectrum of gravitational waves it would produce.
The key parameters we will investigate are the latent heat, the phase transition duration, and the bubble wall velocity.
To this end, we have begun non-perturbative lattice analyses of the latent heat, which will reuse some of the ensembles we have presented here, in addition to more calculations with larger $L$ and $N_t$ in order to extrapolate to the $L \to \infty$ thermodynamic limit and the $N_t \to \infty$ continuum limit.
As $N_t$ increases and the lattice spacing at the confinement transition decreases, we will need to work with smaller \am to stay on a line of constant physics with approximately fixed $M_P / M_V$, which will also add to the numerical costs of these larger-volume calculations.
It will also be challenging to establish robust non-perturbative constraints on the phase transition duration and bubble wall velocity for the first-order stealth dark matter transition, but even so our lattice calculations should be able to provide new insight into those quantities.
In parallel, we can explore whether this transition could generate intergalactic magnetic fields~\cite{Ellis:2019tjf}, and it will also be valuable to investigate alternative approaches for analyzing first-order phase transitions, such as density-of-states techniques~\cite{Langfeld:2015fua}.

\section*{Acknowledgements} 
We are grateful for correspondence and useful discussions with Germano Nardini, R.~V.\ Gavai, Ed Hardy, Owe Philipsen, David Weir and other participants in the ECT* workshop ``Interdisciplinary approach to QCD-like composite dark matter''.
Computing support for this work came from the Lawrence Livermore National Laboratory (LLNL) Institutional Computing Grand Challenge program, as well as from the University of Liverpool.
This work was supported in part by the U.S.~Department of Energy (DOE), Office of Science, Office of High Energy Physics, under Award Number {DE-SC0015845} (RCB and CR), Award Number {DE-SC0019061} (GTF), Award Number {DE-SC0011640} (GDK) and Award Number {DE-SC0010005} (AH, ETN and OW). 
KC was supported by DOE Computational Sciences Graduate Fellowship {DE-SC0019323}.
AG was supported by SNSF Grant Number {200021\_175761}.
ER was supported by a RIKEN SPDR fellowship.
DS was supported by UK Research and Innovation Future Leader Fellowship {MR/S015418/1}.
PV acknowledges the support of the DOE under contract {DE-AC52-07NA27344} (LLNL).
Argonne National Laboratory is supported by the DOE under contract {DE-AC02-06CH11357}.

\newpage
\section*{Appendix: Summary of streams}
Table~\ref{tab:ensembles} summarizes the 1,381 finite-temperature streams we have generated for this work.

\begin{table}[!h]
  \centering
  \renewcommand\arraystretch{1.2}  
  \addtolength{\tabcolsep}{3 pt}   
  \begin{tabular}{lrr|c}
    \hline
    \am      & $N_t$ & $L$ & Streams \\
    \hline
             &       & 16  & 35      \\
    0.05     &  8    & 24  & 35      \\
             &       & 32  & 35      \\
    \hline
    0.067    & 12    & 24  & 19      \\
    \hline
             &       &  8  & 35      \\
             &       & 12  & 21      \\
    0.1      &  4    & 16  & 21      \\
             &       & 24  & 21      \\
             &       & 32  & 21      \\
    \hline
    0.1      &  6    & 12  & 22      \\
    \hline
             &       & 16  & 58      \\
    0.1      &  8    & 24  & 53      \\
             &       & 32  & 53      \\
    \hline
    0.1      & 12    & 24  & 22      \\
    \hline
    0.2      &  4    &  8  & 29      \\
    \hline
    0.2      &  6    & 12  & 25      \\
    \hline
             &       & 16  & 53      \\
    0.2      &  8    & 24  & 51      \\
             &       & 32  & 51      \\
    \hline
    0.2      & 12    & 24  & 25      \\
    \hline
             &       & 12  & 27      \\
    0.4      &  6    & 18  & 27      \\
             &       & 24  & 27      \\
    \hline
             &       & 16  & 63      \\
    0.4      &  8    & 24  & 63      \\
             &       & 32  & 63      \\
    \hline
             &       &  8  & 38      \\
    $\infty$ &  4    & 12  & 38      \\
             &       & 16  & 38      \\
    \hline
             &       & 12  & 37      \\
    $\infty$ &  6    & 16  & 37      \\
             &       & 24  & 37      \\
    \hline
             &       & 16  & 55      \\
    $\infty$ &  8    & 24  & 55      \\
             &       & 32  & 55      \\
    \hline
    $\infty$ & 12    & 24  & 36      \\
    \hline
  \end{tabular}
  \caption{\label{tab:ensembles}A summary of the 1,381 finite-temperature streams generated for this work.  In addition to the $\am = \left\{0.05, 0.1, 0.2, 0.4, \infty\right\}$ highlighted in the body of the paper, we generated a small number of $N_t = 12$ ensembles with $\am = 0.067$ in order to check $N_t$ dependence with fixed $\am\!\cdot\!N_t \approx 0.8$.}
\end{table}

\newpage 
\raggedright
\bibliography{SU4_phase_paper}
\end{document}